\documentclass[twocolumn]{aastex63}

\usepackage{multirow}
\usepackage{comment} 
\usepackage{booktabs}
\usepackage{xcolor}
\definecolor{myPurple}{RGB}{128,0,128} 
\usepackage{soul}

\def\tName{COS-2987}
\def\OIII{[O\,{\sc iii}]}
\def\OIII{[O\,{\sc iii}]}
\def\OII{[O\,{\sc ii}]}
\def\HeI{He\,{\sc i}}
\def\HeII{He\,{\sc ii}}
\def\Halpha{H$\alpha$}
\def\Hbeta{H$\beta$}
\def\Hgamma{H$\gamma$}
\def\Hdelta{H$\delta$}
\def\Hepsilon{H$\epsilon$}
\def\NeIII{[Ne\,{\sc iii}]}
\def\NII{[N\,{\sc ii}]}
\def\SII{[S\,{\sc ii}]}
\def\CII{[C\,{\sc ii}]}

\def\ltsima{$\buildrel<\over\sim$}
\def\la{\lower.5ex\hbox{\ltsima}~}
\def\gtsima{$\buildrel>\over\sim$}
\def\ga{\lower.5ex\hbox{\gtsima}~}
\def\deg~{$^{\circ}$}

\submitjournal{ApJ}
\received{2025/Dec/17}
\accepted{??}

\shorttitle{RIOJA. COS-2987}
\shortauthors{Mawatari et al.}

\begin{document}

\title{RIOJA. A Clumpy Galaxy Assembly at Redshift 6.81 Revealed by JWST}

\correspondingauthor{K. Mawatari}
\email{mawatari@aoni.waseda.jp}
\author[0000-0003-4985-0201]{Ken Mawatari}
\affil{Waseda Research Institute for Science and Engineering, Faculty of Science and Engineering, Waseda University, 3-4-1 Okubo, Shinjuku, Tokyo 169-8555, Japan}
\affiliation{Department of Pure and Applied Physics, School of Advanced Science and Engineering, Faculty of Science and Engineering, Waseda University, 3-4-1 Okubo, Shinjuku, Tokyo 169-8555, Japan}
\affiliation{Division of Physics, Faculty of Pure and Applied Sciences, University of Tsukuba,Tsukuba, Ibaraki 305-8571, Japan}
\affiliation{Tomonaga Center for the History of the Universe (TCHoU), Faculty of Pure and Applied Sciences, University of Tsukuba, Tsukuba, Ibaraki 305-8571, Japan}
\author[0000-0001-6820-0015]{Luca Costantin}
\affiliation{Centro de Astrobiolog\'{\i}a (CAB), CSIC-INTA, Ctra. de Ajalvir km 4, Torrej\'on de Ardoz, E-28850, Madrid, Spain}
\author{Mitsutaka Usui}
\affiliation{Division of Physics, Faculty of Pure and Applied Sciences, University of Tsukuba,Tsukuba, Ibaraki 305-8571, Japan}
\author[0000-0002-0898-4038]{Takuya Hashimoto}
\affiliation{Division of Physics, Faculty of Pure and Applied Sciences, University of Tsukuba,Tsukuba, Ibaraki 305-8571, Japan}
\affiliation{Tomonaga Center for the History of the Universe (TCHoU), Faculty of Pure and Applied Sciences, University of Tsukuba, Tsukuba, Ibaraki 305-8571, Japan}
\author[0000-0002-7093-1877]{Javier \'Alvarez-M\'arquez}
\affiliation{Centro de Astrobiolog\'{\i}a (CAB), CSIC-INTA, Ctra. de Ajalvir km 4, Torrej\'on de Ardoz, E-28850, Madrid, Spain}
\author[0000-0001-6958-7856]{Yuma Sugahara}
\affiliation{Waseda Research Institute for Science and Engineering, Faculty of Science and Engineering, Waseda University, 3-4-1 Okubo, Shinjuku, Tokyo 169-8555, Japan}
\affiliation{Department of Pure and Applied Physics, School of Advanced Science and Engineering, Faculty of Science and Engineering, Waseda University, 3-4-1 Okubo, Shinjuku, Tokyo 169-8555, Japan}
\author[0000-0002-9090-4227]{Luis Colina}
\affiliation{Centro de Astrobiolog\'{\i}a (CAB), CSIC-INTA, Ctra. de Ajalvir km 4, Torrej\'on de Ardoz, E-28850, Madrid, Spain}
\author[0000-0002-7779-8677]{Akio K. Inoue}
\affiliation{Waseda Research Institute for Science and Engineering, Faculty of Science and Engineering, Waseda University, 3-4-1 Okubo, Shinjuku, Tokyo 169-8555, Japan}
\affiliation{Department of Pure and Applied Physics, School of Advanced Science and Engineering, Faculty of Science and Engineering, Waseda University, 3-4-1 Okubo, Shinjuku, Tokyo 169-8555, Japan}
\author{Wataru Osone}
\affiliation{Division of Physics, Faculty of Pure and Applied Sciences, University of Tsukuba,Tsukuba, Ibaraki 305-8571, Japan}
\author[0000-0001-7997-1640]{Santiago Arribas}
\affiliation{Centro de Astrobiolog\'{\i}a (CAB), CSIC-INTA, Ctra. de Ajalvir km 4, Torrej\'on de Ardoz, E-28850, Madrid, Spain}
\author[0000-0001-8442-1846]{Rui Marques-Chaves}
\affiliation{Geneva Observatory, Department of Astronomy, University of Geneva, Chemin Pegasi 51, CH-1290 Versoix, Switzerland}
\author[0000-0002-0984-7713]{Yurina Nakazato}
\affiliation{Center for Computational Astrophysics, Flatiron Institute, 162 5th Avenue, New York, NY, 10010, USA}
\affiliation{Department of Physics, The University of Tokyo, 7-3-1 Hongo, Bunkyo, Tokyo 113-0033, Japan}
\author[0000-0001-8083-5814]{Masato Hagimoto}
\affiliation{Department of Physics, Graduate School of Science, Nagoya University, Nagoya, Aichi 464-8602, Japan}
\author{Takeshi Hashigaya}
\affiliation{Department of Astronomy, Kyoto University Sakyo-ku, Kyoto 606-8502, Japan}
\author[0000-0002-8680-248X]{Daniel Ceverino}
\affiliation{Universidad Autonoma de Madrid, Ciudad Universitaria de Cantoblanco, E-28049 Madrid, Spain}
\affiliation{CIAFF, Facultad de Ciencias, Universidad Autonoma de Madrid, E-28049 Madrid, Spain}
\author[0000-0001-7925-238X]{Naoki Yoshida}
\affiliation{Department of Physics, The University of Tokyo, 7-3-1 Hongo, Bunkyo, Tokyo 113-0033, Japan}
\affiliation{Kavli Institute for the Physics and Mathematics of the Universe (WPI), UT Institute for Advanced Study, The University of Tokyo, Kashiwa, Chiba 277-8583, Japan}
\affiliation{Research Center for the Early Universe, School of Science, The University of Tokyo, 7-3-1 Hongo, Bunkyo, Tokyo 113-0033, Japan}
\author[0000-0002-5268-2221]{Tom J. L. C. Bakx}
\affiliation{Department of Space, Earth, \& Environment, Chalmers University of Technology, Chalmersplatsen 4 412 96 Gothenburg, Sweden}
\affiliation{Department of Physics, Graduate School of Science, Nagoya University, Nagoya, Aichi 464-8602, Japan}
\affiliation{National Astronomical Observatory of Japan, 2-21-1, Osawa, Mitaka, Tokyo, Japan}

\author[0000-0001-7440-8832]{Yoshinobu Fudamoto}
\affiliation{Center for Frontier Science, Chiba University, 1-33 Yayoi-cho, Inage-ku, Chiba 263-8522, Japan}
\author[0000-0003-2119-277X]{Alejandro Crespo G\'omez}
\affiliation{Space Telescope Science Institute (STScI), 3700 San Martin Drive, Baltimore, MD 21218, USA}
\author[0000-0003-3278-2484]{Hiroshi Matsuo}
\affiliation{National Astronomical Observatory of Japan, 2-21-1, Osawa, Mitaka, Tokyo, Japan}
\affiliation{Graduate University for Advanced Studies (SOKENDAI), 2-21-1 Osawa, Mitaka, Tokyo 181-8588, Japan}
\author[0000-0002-4005-9619]{Miguel Pereira-Santaella}
\affiliation{Instituto de F\'isica Fundamental (IFF), CSIC, Serrano 123, E-28006, Madrid, Spain}
\author[0009-0005-5448-5239]{Carmen Blanco-Prieto}
\affiliation{Centro de Astrobiolog\'{\i}a (CAB), CSIC-INTA, Ctra. de Ajalvir km 4, Torrej\'on de Ardoz, E-28850, Madrid, Spain}
\author{Yi W. Ren}
\affiliation{Department of Pure and Applied Physics, School of Advanced Science and Engineering, Faculty of Science and Engineering, Waseda University, 3-4-1 Okubo, Shinjuku, Tokyo 169-8555, Japan}
\author[0000-0003-4807-8117]{Yoichi Tamura}
\affiliation{Department of Physics, Graduate School of Science, Nagoya University, Nagoya, Aichi 464-8602, Japan}
%




\begin{abstract}

Spatially resolved multi-wavelength analysis is essential to study galaxy formation and evolution. A UV-bright galaxy COS-2987030247 at $z = 6.81$ is one of the Rosetta Stones in the epoch of reionization for which JWST NIRSpec Integral Field Spectroscopy, NIRCam imaging, and ALMA data are available thanks to the RIOJA program. We identified the rest-frame optical emission lines from the ionized hydrogen, oxygen, and neon gas. The \OIII\,5008\AA\ line emission and the NIRCam images show a complex kinematical and morphological structure where two bright main and three faint clumps are identified in a 10 kpc extent.
The system is not classified as a purely rotation-dominated disk. The multiple clumps are instead consistent with a merger-related origin, including either distinct galaxies in interaction or star-forming clumps formed through tidal gas compression during a merger.
The spatially resolved emission line fluxes show that dust attenuation, metal enrichment, and ionization parameter are preferentially enhanced in the star formation peaks. Our SED fitting suggests that the main clumps are in a moderately dust-attenuated star forming phase ($A_{\rm V} = 0.2$--$0.3$ and SFR(H$\alpha$) $\sim 10$\,M$_\odot$\,yr$^{-1}$) with almost zero escape fraction of ionizing photons. In contrast, the sub-clumps are dust-free and lying on or below the main sequence of star-forming galaxies. These sub-clumps may work as a perturber that triggers the clumpy starburst in the surrounding gas through the merger event. 

\end{abstract}



\section{Introduction} \label{sec:intro}

Understanding how galaxies form and evolve requires spatially and kinematically resolved views of their internal structure. Integral field spectroscopy (IFS) is a powerful technique for probing internal structures such as star-forming regions and the interstellar medium (ISM). Ground-based IFS instruments, assisted by adaptive optics (AO), revealed the morphological and kinematic evolution of galaxies up to redshift $z \sim 2$--3 \citep[e.g.,][]{Genzel+06,Genzel+08,ForsterSchreiber+09,Gnerucci+11,Contini+12,Law+09,Wisnioski+15,Stott+16,ForsterSchreiber+18}. With the advent of the {\it James Webb Space Telescope} (JWST; \citealt{Gardner+23}) and the Atacama Large Millimeter/submillimeter Array (ALMA;\citealt{WoottenThompson09}), it has become possible to study galaxies at $z \gtrsim 6$ with unprecedented spatial and spectral resolution, enabling rapid progress in characterizing star formation, feedback, and chemical enrichment processes in the reionization era \citep[e.g.,][]{Smit+18,Parlanti+23,Fudamoto+22,Marconcini+24,Xu+24,Jones+24a,Arribas+24,Telikova+25,Gimenez-Arteaga+24,Jones+25}. 

Previous studies have shown that galaxies at high redshift exhibit irregular morphologies and disturbed kinematics, such as clumpy structures and asymmetric velocity fields \citep[e.g.,][]{Guo+15,Wisnioski+15,Simons+16,Shibuya+16,Rodrigues+17,Chen+23,Sugahara+25,Scholtz+25,Shibuya+25,Harikane+25a}. The increasing prevalence of such irregular systems toward earlier cosmic times is broadly consistent with the rising galaxy major merger rate, which is predicted to scale as $\propto (1+z)^{2-3}$ \citep[e.g.,][]{FakhouriMa08,Conselice+09,Rodriguez-Gomez+15,Duan+25}. Mergers are known to disrupt both morphology and kinematics, often enhancing gas accretion and triggering bursts of star formation (\citealt{MihosHernquist94,MihosHernquist96,BarnesHernquist96,DiMatteo+07}; but see also \citealt{Fensch+17}). Recent JWST observations have further revealed extreme examples of merger-driven systems at high redshift, including multi-galaxy assemblies with numerous clumps and extended ionized gas structures \citep[e.g.,][]{Hu+25}. Beyond mergers, theoretical studies have shown that gas-rich and highly turbulent disks at high redshift are inherently susceptible to gravitational instability, which can give rise to massive star-forming clumps
\citep[e.g.,][]{KrumholzBurkert10,Romeo+10,Forbes+14,RomeoAgertz14,InoueS+16}. Simulations also have suggested that cold gas inflows along cosmic filaments, referred to as ``cold streams'', can drive violent disk instabilities, leading to the formation of massive star-forming clumps within gas-rich disks \citep[e.g.,][]{Dekel+09,Ceverino+10,DekelKrumholz13,Mandelker+17}. These scenarios provide theoretical support for the prevalence of clumpy galaxies at high redshift.

At the same time, recent ALMA and JWST observations have revealed massive, rotation-dominated disk galaxies even at $z > 3$, with dynamical properties comparable to those of nearby spiral galaxies (\citealt{Rizzo+20, Neeleman+20,TsukuiIguchi21,Costantin+23,Umehata+25}; see also \citealt{Parlanti+23,Parlanti+24,Fujimoto+25}). This growing diversity of morpho-kinematic structures in the early Universe underscores the need for spatially and spectrally resolved studies of galaxies at $z > 6$. Such observations are essential for understanding the physical mechanisms that regulate star formation modes and for revealing the role of galaxies in the cosmic reionization.

This study is part of the RIOJA project: ``the Reionization and the ISM/Stellar Origins with JWST and ALMA'' \citep[JWST GO1 PID1840; PIs: J. \'{A}lvarez-M\'{a}rquez and T. Hashimoto;][]{Alvarez-Marquez+21,Hashimoto+23a,Sugahara+25,Ren+25}. The project aims to investigate galaxy evolution during the cosmic reionization through joint analyses of high-resolution IFS and imaging data. A key feature of RIOJA is the use of ALMA-detected \OIII\,88$\mu$m emitters as anchor targets, which enables detailed, multi-phase ISM studies when combined with JWST/NIRSpec IFS and NIRCam imaging data.

One of the RIOJA program targets, COS-2987030247 (hereafter \tName), located at (R.A., Decl.) = ($10^{\rm h}00^{\rm m}29.87^{\rm s}$, $+2^{\circ}13^{\prime}02\farcs47$), has been the subject of intensive study since the pre-JWST era. This galaxy was initially identified by \citet{Smit+15} as a $z \sim 6.8$ Lyman Break Galaxy (LBG) with an extreme Spitzer/IRAC \citep{Fazio+04} $[3.6] - [4.5]$ color, indicative of strong \OIII\,5008\AA$+$4960\AA$+$H$\beta$ emission. Its well-constrained photometric redshift and large \OIII\,5008\AA\ equivalent width (EW) made it an excellent candidate for follow-up studies with ALMA. \citet{Smit+18} subsequently detected luminous \CII\,158$\mu$m emission from \tName\ and identified a clear velocity gradient, suggesting a possible rotating disk. \citet{Laporte+17} conducted deep near-infrared spectroscopy and detected Ly$\alpha$ emission, suggesting the presence of an early ionized bubble surrounding the galaxy. More recently, \citet{Witstok+22} and \citet{Posses+23} presented spatially resolved ALMA observations of \OIII\,88$\mu$m and \CII\,158$\mu$m. \citet{Witstok+22} found that the \OIII\,88$\mu$m emission is compact and aligns with the UV morphology, whereas \CII\,158$\mu$m appears more extended, reflecting a multiphase ISM and possibly enhanced $\alpha$-element abundances. \citet{Posses+23} modeled the \CII\,158$\mu$m kinematics using \texttt{$^{\rm 3D}$BAROLO} \citep{DiTeodoroFraternali15} and found that \tName\ is consistent with a rotating, low-dispersion disk viewed nearly face-on, although the presence of faint extended structures leaves open the possibility of a compact merger. These pre-JWST studies established \tName\ as a benchmark system for probing the ISM and dynamical conditions in the epoch of reionization.

With the advent of JWST, whose unprecedented sensitivity and spatial resolution enables detailed mapping of the rest-frame optical emission at high redshift, the disk-like interpretation of COS-2987 has begun to shift. High-resolution NIRCam imaging reveals that COS-2987 is not a smooth, monolithic structure, but instead comprises multiple clumps, suggesting a more complex dynamical state. 
\citet{Harikane+25a} analyzed a large sample of spectroscopically confirmed galaxies at $z = 7$–$14$, including \tName, and found that $\sim 70$\,\% of UV-luminous galaxies at $z \sim 7$ exhibit clumpy morphologies indicative of merger-induced starburst activity. \citet{Harikane+25a} performed SED fitting for individual clumps in \tName, but their analysis based solely on imaging data. In contrast, our study combines JWST/NIRCam imaging with JWST/NIRSpec IFS and ALMA spectroscopy, offering spatially resolved line diagnostics as well as comprehensive SED fitting. This approach allows for a more accurate derivation of key physical quantities for each clump, and more importantly, facilitates the construction of a coherent physical picture of the galaxy at reionization era. 

This paper is organized as follows. Section~\ref{sec:JWSTobs} describes the JWST and ALMA observations and data reduction. Section~\ref{sec:LineAna} presents spatially resolved emission-line analyses based on \OIII\,5008\AA, including kinematic and morphological properties. Section~\ref{sec:multiwaveana} focuses on SED analyses of individual clumps using multi-wavelength photometry and spectroscopy. In Section~\ref{sec:disc_view}, we combine all the observational results and physical diagnostics to construct a comprehensive physical picture of COS-2987, offering an integrated interpretation of its internal structure, evolutionary state, and the possible physical origin of its clumpy morphology. This section also serves as the conclusion of the paper. Throughout this paper, we adopt a cosmology with $H_{0}=70$\,km\,s$^{-1}$\,Mpc$^{-1}$, $\Omega_{M}=0.3$, and $\Omega_{\Lambda}=0.7$. We assumed a Chabrier initial mass function (IMF; \citealt{Chabrier03}).

\section{Observation and Calibration} \label{sec:JWSTobs}

\subsection{JWST NIRCam}  \label{sec:NIRCAMobs}

\begin{table*}[]
\begin{center}
\caption{Summary of NIRCam observations} \label{tb:NIRCam_obs}
\begin{tabular}{cccccc}
\hline
\hline
 Filter & Program ID\tablenotemark{\rm a} & Exposure Time & PSF FWHM & Depth in $2 \times$ PSF FWHM\tablenotemark{\rm b} & Depth in $0\farcs5$\tablenotemark{\rm b} \\
  &  & (hours) & ($^{\prime\prime}$) & (nJy) & (nJy)  \\
\hline
F115W & 1837, 1840 & 0.92 & 0.06 & 7.0 & 48.0 \\
F150W & 1837, 1840 & 0.85 & 0.06 & 6.7 & 40.3 \\
F200W & 1837, 1840 & 0.90 & 0.07 & 6.0 & 34.0 \\
F250M & 1840 & 0.23 & 0.12 & 62.2 & 169.7 \\
F277W & 1837 & 0.70 & 0.12 & 8.8 & 22.3 \\
F335M & 1840 & 0.20 & 0.14 & 44.5 & 92.2 \\
F356W & 1837 & 0.70 & 0.14 & 9.8 & 19.7 \\
F444W & 1837, 1840 & 0.85 & 0.16 & 15.2 & 27.7 \\
\hline
\end{tabular}
\end{center}
\tablenotetext{\rm a}{JWST program ID. \#1837: PRIMER \citep{Dunlop+21}; \#1840: RIOJA \citep{Alvarez-Marquez+21}.}
\tablenotetext{\rm b}{The $5\sigma$ limiting flux with $2\times$ PSF FWHM and $0\farcs5$ diameter apertures were estimated from random aperture photometry on blank sky region.}
\end{table*}

The JWST data were obtained as part of the RIOJA project. The specific observations can be accessed via the Mikulski Archive for Space Telescopes (MAST) at the Space Telescope Science Institute: \dataset[doi:10.17909/c5cd-b767]{https://doi.org/10.17909/c5cd-b767}.

We used NIRCam data taken in the two JWST GO1 programs \#1840 (RIOJA) and \#1837 (PRIMER; PI: J.~Dunlop; \citealt{Dunlop+21}). In total, six broad-bands and two medium-bands data are used in this work. The observational details are summarized in Table~\ref{tb:NIRCam_obs}. All raw data were reduced using the JWST pipeline version 1.17.1 (\citealt{Bushouse+22}) under CRDS context \texttt{jwst\_1321.pmap}. We also applied custom procedures which are the same as those in \citet{Hashimoto+23a} and \citet{Sugahara+25}: 1) snowball and wisps correction as described in \citet{Bagley+23}, and 2) a background homogenization before stacking all exposures following \citet{PerezGonzalez+23}. For further improving the overall background characterization, we applied \texttt{NoiseChisel} \citep{AkhlaghiIchikawa15,Golini+24} which creates suitable masks of every sources covering diffuse emission from galaxy outskirts and bright star spikes \citep{PerezGonzalez+25}. The final mosaics are all drizzled with a pixel scale of $0\farcs03$ pix$^{-1}$. 

The Point Spread Function (PSFs) were derived by stacking $N = 20$ -- $70$ point sources in each band image. The corresponding full widths at half maximum (FWHMs) range from $0\farcs06$ to $0\farcs16$ (Table~\ref{tb:NIRCam_obs}). We cutout $\sim 20'' \times 40''$ regions of uniform depths around \tName\ for the derivation of the limiting flux. We measured noise fluctuation by $N = 500$ random aperture photometry on blank sky region. To match aperture sizes used for actual photometry to multiple clumps in \tName\ (Section~\ref{sec:MultiPhot}), we repeated the random aperture photometry changing the aperture diameter from $0\farcs1$ to $2\farcs0$. The $5\,\sigma$ limiting fluxes in the cases with aperture diameters $= 0\farcs5$ and $2 \times$ PSF FWHM are shown in Table~\ref{tb:NIRCam_obs}. 

We calibrated the astrometry of the NIRCam images using the GAIA DR3 stars \citep{Gaia2016}. We carefully selected $N = 4$ -- $15$ GAIA stars that are not saturated in the NIRCam images. After the absolute calibration using the GAIA stars, we further corrected for relative positional offset among the filters, where the point sources in F444W were used as the WCS reference. The WCS solutions were estimated using \texttt{IRAF} \citep{Tody86,Tody93}. We eventually achieved astrometric accuracy of $\lesssim 0\farcs02$ against both the GAIA stars and the F444W reference objects.

\subsection{JWST NIRSpec IFS}  \label{sec:NIRSPECobs}

As part of the RIOJA project, we conducted NIRSpec IFS observations of \tName\ on 26 May 2023. The observations were taken with a grating/filter pair of G395H/F290LP that produced a cube with a spectral resolution of $R \sim 2700$ in a wavelength range of $2.87$ -- $5.27$\,\micron. Four dithers were set with a medium size cycling dither pattern. The total on-source integration time was 2626 seconds. 

We reduced the raw data with the JWST pipeline version 1.14.0 under CRDS context \texttt{jwst$\_$1223.pmap}. Following the data reduction process used by GTO programs (e.g., \citealt{Marshall+23, Ubler+23, Perna+23}) and our previous RIOJA work \citep{Hashimoto+23a}, we applied some modifications to the pipeline including (1) the removal of the $1/f$ noise (c.f., \citealt{Bagley+23}), (2) rejection of the bad pixels and cosmic rays by sigma-clipping, and (3) removal of a median background in the calibrated images. The data cube was sampled with a pixel size of $0\farcs05$. A scale of the pixel along the spectral direction (spectral element) is $6.65$\,\AA\ corresponding to 40 -- 70\,km\,s$^{-1}$ which is high enough to sample the spectral resolution of 80 -- 150\,km\,s$^{-1}$. The effective area is about $3\farcs8 \times 3\farcs8$.  

To set the accurate astrometry, we cross-correlated the NIRSpec \OIII\,5008\,\AA\ integrated intensity map with the NIRCam F356W image using peak positions of three significantly detected objects (mN, mS, and sN1 in Section~\ref{sec:MultiPhot}). The reference NIRCam F356W image is significantly contaminated by the \OIII\,5008\,\AA\ emission by $\sim 32$\,\%. The astrometric accuracy is $< 0\farcs01$ against the F356W image. 

We also used NIRSpec IFS data of the A3V standard star ``1808347'' (2MASS J18083474+6927286) to evaluate the PSF at each spectral element in the 3D cube. The data were taken in the commissioning program (PID1128; PI: N. L\"{u}tzgendorf) with the same filter and grating as ours. We reduced the standard star data with the same JWST pipeline version and CRDS context as for \tName.

\subsection{ALMA} \label{sec:ALMAdata}

Archival ALMA data in Bands\,5 \citep{Belitsky+18}, 6 \citep{Ediss+04}, and 8 \citep{Sekimoto+08} are available for \tName. These data were taken aiming at detection of the \NII\,205$\mu$m, \CII\,158$\mu$m, \OIII\,88$\mu$m lines and dust continuum. 

We re-analyzed the Band\,8 data because we compare morphology and line profile between ALMA \OIII\,88$\mu$m and NIRSpec \OIII\,5008\AA\ in a companion paper \citep{Usui+25}. The raw data are same as \citet{Witstok+22}, obtained in ALMA Cycle 6 program (PID: 2018.1.00429.S; PI: R. Smit). The data reduction was performed with the Common Astronomy Software Application (\texttt{CASA}; \citealt{McMullin+07,CASATeam+22}) version 5.6. We produced line cube and continuum image using \texttt{tclean} task with the natural weighting to maximize sensitivity for point-sources. No $uv$-tapering was adopted. In the line cube construction, we set the channel width as 30\,km\,s$^{-1}$. The \OIII\,88$\mu$m line emission is clearly detected, whose redshift and morophology are consistent with those of \citet{Witstok+22}. Typical noise root-mean-square (rms) around the \OIII\,88$\mu$m line is $0.576$\,mJy per 30\,km\,s$^{-1}$ bin. We integrated the cube data over a velocity range of $480$\,km\,s$^{-1}$ centered at the \OIII\,88$\mu$m line peak. In this \OIII\,88$\mu$m line integrated intensity map, the beam size is $0\farcs80 \times 0\farcs61$ with a position angle of $86^{\circ}$, and noise rms is $82.6$\,mJy\,beam$^{-1}$\,km\,s$^{-1}$. In the continuum image construction, we excluded the spectral window that includes the redshifted \OIII\,88$\mu$m line. The beam size of the continuum map is $0\farcs82 \times 0\farcs62$ with a position angle of $86^{\circ}$, and the noise rms is 70.4\,$\mu$Jy\,beam$^{-1}$. \tName\ is not detected in this continuum image, consistent with \citet{Witstok+22}. 

In Band\,5, we used measurements of \citet{Witstok+22}. They reduced the data taken in ALMA Cycle 6 program (PID: 2018.1.01551.S; PI: R. Smit), resulting in the beam size of $0\farcs8 \times 0\farcs7$ and the noise rms of $0.166$\,mJy\,beam$^{-1}$ per 48\,km\,s$^{-1}$ bin. They produced the continuum image by averaging over all available spectral channels with natural weighting. The noise rms in the continuum image is $6.47\,\mu$Jy\,beam$^{-1}$. They detect neither \NII\,205$\mu$m emission line nor dust continuum for \tName.  

In Band\,6, we used measurements of \citet{Posses+23} who combined data taken in two ALMA programs (PID: 2015.1.01111.S; PI: R. Smit and PID: 2018.1.01359.S; PI: M. Aravena). They achieved high angular resolution, beam size $= 0\farcs44 \times 0\farcs35$. The noise rms is $0.134$\,mJy\,beam$^{-1}$ per 30\,km\,s$^{-1}$ bin and $6\,\mu$Jy\,beam$^{-1}$ for the line cube and continuum image, respectively. They detected the \CII\,158$\mu$m line emission from \tName\ that is spatially resolved thanks to the high angular resolution. On the other hand, dust continuum is not detected.

\section{Emission Line Analyses in the NIRSpec IFS Data}\label{sec:LineAna}

\subsection{Emission Line Search} \label{sec:LineSearch1}

\begin{figure*}
    \centering
    \includegraphics[width=1.0\linewidth, angle=0]{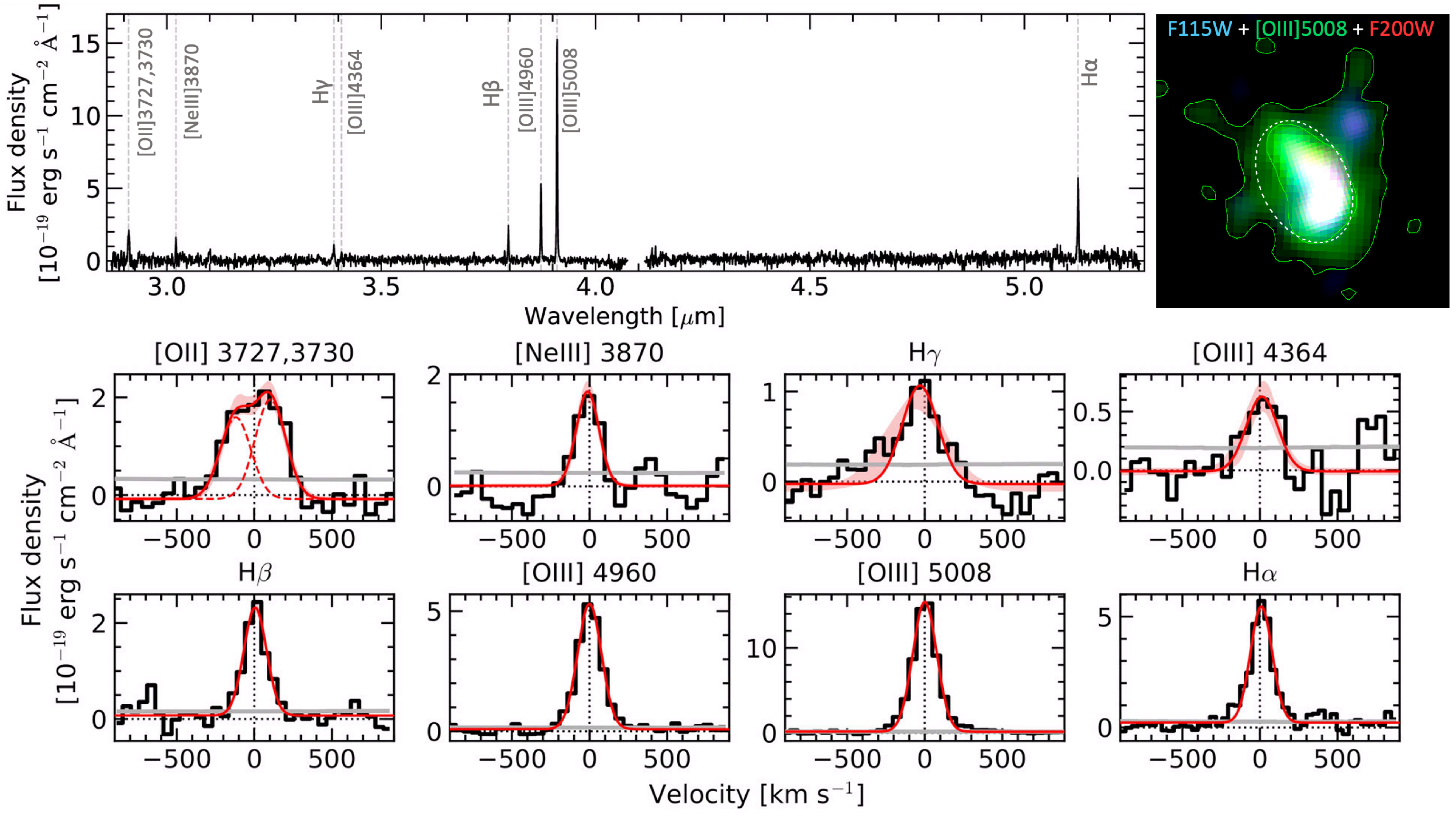}
    \caption{(Top panel) One-dimensional spectrum of \tName\ extracted from the NIRSpec IFS data cube. Vertical dashed lines indicate the detected emission lines. (Right panel) The aperture used for extraction of the spectrum is shown by a dashed ellipse which is superposed on the F115W (blue) $+$ \OIII\,5008\AA\ (green) $+$ F200W (red) pseude color image. These three images were smoothed so that their PSFs are matched (section~\ref{sec:MultiPhot}). Contours show the \OIII\,5008\AA\ integrated intensity map at $2 \times$ and $10 \times \sigma$. Panel size is $2'' \times 2''$ corresponding to 11.6\,kpc $\times$ 11.6\,kpc at $z = 6.81$. (Bottom panels) Zoom-in spectra of the detected emission lines, where the horizontal axis is the velocity relative to the best-fit redshift to \OIII\,5008\AA, $z = 6.80731$. In each panel, the black solid histogram and gray line indicate the observed and $1\sigma$ error spectra, respectively. The red solid line and red shade show the best-fit Gaussian profile and $68$\,\% confidential interval from the MC iteration (see text). For the \OII\,3727,3730\AA\ doublet, the best-fit models for individual lines are shown by the red dashed lines. }
    \label{fig:1DwholeSpec}
\end{figure*}

In the reduced NIRSpec IFS data cube, we identified bright \OIII\,5008\AA\ line emission at $z \approx 6.81$. To accurately determine the redshift and search for other emission lines, we extracted one-dimensional spectrum using \texttt{photutils} \citep{Bradley+23} with an elliptical aperture whose center position, semi-major axis, semi-minor axis, and position angle are (R.A., Decl.) = ($10^{\rm h}00^{\rm m}29.870^{\rm s}$, $+2^{\circ}13^{\prime}02\farcs266$), $0\farcs45$, $0\farcs30$, and 27.2\,degrees, respectively. This aperture roughly covers the two brightest emission peaks (mN and mS in Section~\ref{sec:kinematics}). We set this relatively small aperture to optimize signal-to-noise ratio (SNR) and used it only for the emission line search. The aperture and extracted one-dimensional spectrum are shown in Figure~\ref{fig:1DwholeSpec}. We estimated the associated error spectrum by measuring standard deviation of the flux density over a wavelength range within $\pm 0.05\,\mu$m centered on each spectral element.  

We searched for the following emission lines\footnote{The rest-frame wavelengths of the emission lines are extracted from Linetools \citep{Prochaska+17} and Atomic Sprctra Database \citep{NIST_ASD} in vacuum.}: \OII\,3727,3730\AA\ doublet, H9, \NeIII\,3870\AA, H8, \HeI\,3966\AA, \Hepsilon, \HeI\,4027\AA, \Hdelta, \Hgamma, \OIII\,4364\AA, \HeI\,4473\AA, \HeII\,4687\AA, \Hbeta, \HeI\,4923\AA, \OIII\,4960\AA, \OIII\,5008\AA, \HeI\,5017\AA, \HeI\,5877\AA, \NII\,6550\AA, \Halpha, \NII\,6585\AA, \HeI\,6680\AA, \SII\,6718\AA, and \SII\,6733\AA. 
Gaussian profile fits were applied to the spectra cutout around expected wavelengths of the above lines at $z \approx 6.81$. Since the purpose of this fitting is to evaluate significance of the line detection, we avoid fixing the redshift and line width to those of the brightest \OIII\,5008\AA. For the \OII\,3727,3730\AA\ doublet we fit the cutout spectrum with double Gaussian profiles whose separation was fixed and widths were set to be the same. The best-fit Gaussian profiles to the bright lines are shown in Figure~\ref{fig:1DwholeSpec}. The horizontal axis is the velocity relative to the best-fit redshift to \OIII\,5008\AA, $z = 6.80731 \pm 0.00002$. To evaluate the fitting uncertainty, we followed a Monte-Carlo (MC) manner. We repeatedly performed the Gaussian profile fitting to 1,000 mock spectra that were generated by perturbing the observed spectrum. Fluctuation of the best-fit parameters from the 1,000 iterations defines $68$\,\% confidential intervals ($1\,\sigma$ errors) of the line properties such as redshift, flux, and line width. 

We calculated two types of SNRs for each line: peak flux density SNR and integrated flux SNR. We also produced line intensity maps for visual inspection by integrating the data cube over a wavelength range corresponding to $3 \times$ FWHM around the line peak. We set detection criteria as follows: 1) peak SNR $\geq 2$, 2) integrated SNR $\geq 3$, and 3) good in visual inspection of the integrated intensity map as well as the cube data. If a given line passes all or two of the three criteria, the line is regarded as detection. Otherwise, we consider the line as nondetection. While only the second criterion on the integrated SNR may be enough as shown in some previous works using JWST NIRSpec MSA data (e.g., \citealt{Laseter+24,Morishita+24b}), the IFS data allow us to more robustly determine whether a given line is detected by the third criterion. In \tName, the detected lines are \OII\,3727,3730\AA\ doublet, \NeIII\,3870\AA, \Hgamma, \OIII\,4364\AA, \Hbeta, \OIII\,4960\AA, \OIII\,5008\AA, and \Halpha. While the \OIII\,4364\AA\ line appears faint in the integrated intensity map, its peak and integrated SNRs are as high as 3.3 and 4.2, respectively. These SNR levels are usually considered detection \citep{Nakajima+23,Morishita+24b,Laseter+24}.

\subsection{Morphology and Kinematics in [O\,{\sc iii}]\,5008\AA} \label{sec:kinematics}

\begin{figure*}
    \centering
    \includegraphics[width=1.0\linewidth, angle=0]{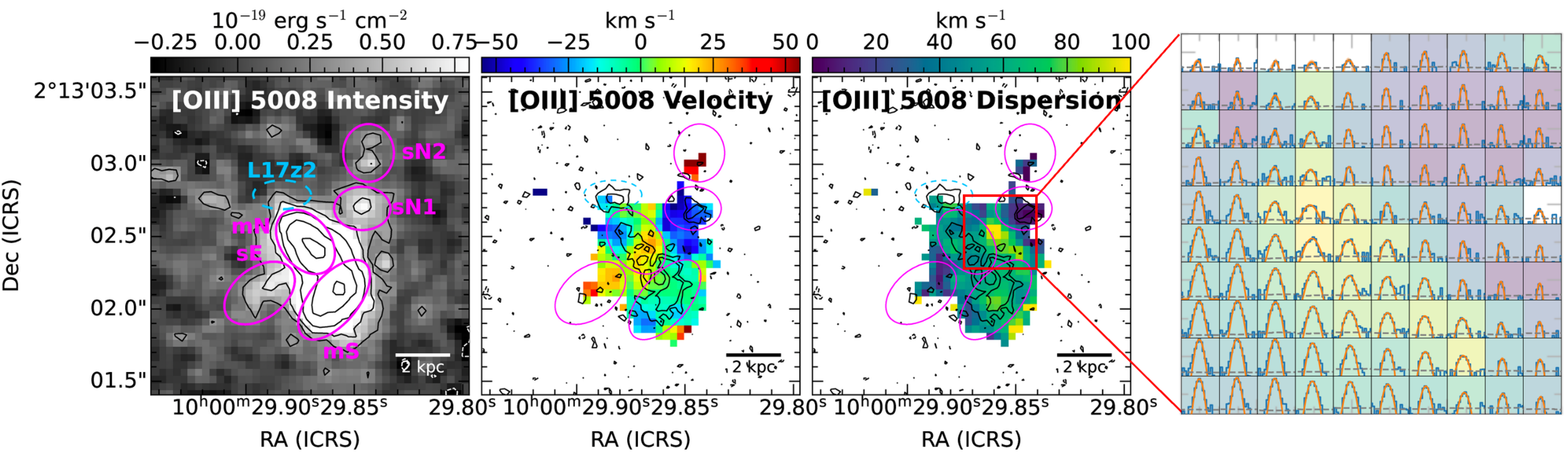}
    \caption{The \OIII\,5008\AA\ line integrated intensity (left), velocity (middle), and velocity dispersion (right) maps. Each panel has $2\farcs2 \times 2\farcs2$ in size. Ellipses correspond to the photometric apertures for the multiple clumps (section~\ref{sec:MultiPhot}). Contours in the integrated intensity map are $-2$ (dashed), 2, 4, 8, 16, and $32 \times \sigma$ (solid) of the line brightness. Contours in the velocity and velocity dispersion maps show the F115W surface brightness profile before the PSF matching at $-2$ (dashed), 2, and $4 \times \sigma$ (solid) levels. The F115W surface brightness is sensitive to 100\,Myr time-scale star formation. The rightmost panel is also the velocity dispersion map but zoomed around the highest velocity dispersion region between mN and sN1. We also superposed the observed (blue), error (grey dashed) spectra, and the best-fit Gaussian profile (orange) on each pixel. }
    \label{fig:mommaps}
\end{figure*}

\begin{figure}
    \centering
    \includegraphics[width=1.0\linewidth, angle=0]{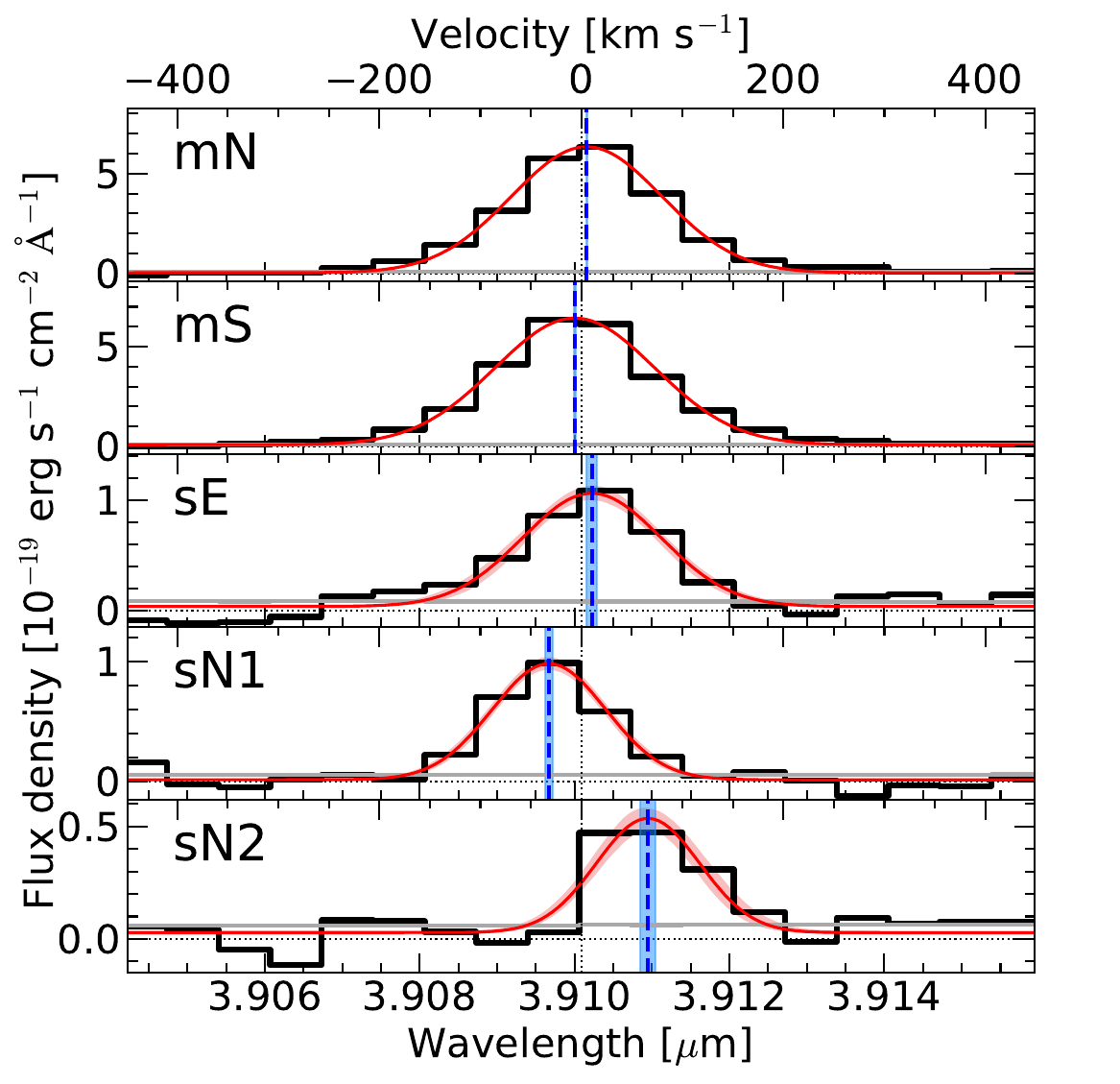}
    \caption{The \OIII\,5008\AA\ line profiles of the five clumps. The velocity axis is centered at $z = 6.80731$. In each panel, the black solid histogram and gray line indicate the observed and $1\sigma$ error spectra, respectively. The red solid line and red shade show the best-fit Gaussian profile and $68$\,\% confidential interval from the MC iteration. The blue vertical dashed line and shade indicate the best-fit peak velocity and uncertainty of each clump. }
    \label{fig:OIII5008specFits}
\end{figure}

We investigated the morphology and kinematics of \tName\ using \OIII\,5008\AA\ that has the highest SNR among the lines. Figure~\ref{fig:mommaps} shows the integrated intensity, velocity, and velocity dispersion maps of \OIII\,5008\AA. The line intensity map was produced following the same procedure described in Section~\ref{sec:LineSearch1}. To produce the velocity and velocity dispersion maps, we performed a Gaussian profile fit for the spectrum at each spatial pixel in the cube data (commonly used technique; e.g., \citealt{Smit+18,Ubler+19,Tokuoka+22,Perna+23}). The fitting procedures are the same as those adopted for the integrated spectrum (Section~\ref{sec:LineSearch1}). The velocity dispersion was corrected for the instrumental spectral resolution \citep{Jakobsen+22}. We masked spatial pixels where the fitting significance was less than $5\,\sigma$. The fitting significance was evaluated from the $\chi^2$ distribution \citep{Tokuoka+22}. 

The line integrated intensity map (the left panel in Figure~\ref{fig:mommaps}) shows a complex morphology. On this map, we visually identified five spatially distinct structures showing either a clear local intensity peak or a linear, tail-like extension. The \OIII\,5008\AA\ flux of \tName\ mainly comes from two main north and south peaks which we refer to as ``mN'' and ``mS'' clumps. In the northwest direction from these central clumps, there are two other fainter peaks. We refer to these substructures as ``sN1'' and ``sN2'' clumps. We identified a tail-like substructure ``sE'' extending southeast of the mN clump. Throughout this paper, we use the term ``clump'' in an observational sense to refer to spatially distinct regions identified in the NIRSpec IFS and NIRCam data\footnote{The term ``clump'' in this paper does not necessarily correspond to a stellar system typically assumed to form within a host galaxy.}.

In the velocity map (middle panel in Figure~\ref{fig:mommaps}), the five clumps have different velocities. 
We quantitatively evaluated the velocity differences by the clumps (Section~\ref{sec:MultiPhot}) and along a given axis (Appendix~\ref{ap:kinematics}). Figure~\ref{fig:OIII5008specFits} shows \OIII\,5008\AA\ line profiles of the five clumps, where the peak velocities and uncertainties relative to $z = 6.80731$ are shown by vertical dashed lines and shades. These velocity offsets are also shown in Table~\ref{tb:fluxAll}. Compared to mN, sN1 is blueshifted by $-37$\,km\,s$^{-1}$ and sN2 is redshifted by $+61$\,km\,s$^{-1}$. The positive velocity of sE relative to mN, $\Delta v = 5$\,km\,s$^{-1}$, is not significant due to the large uncertainty of the faint sE tail. The velocity difference between mN and mS, $\sim 12$\,km\,s$^{-1}$, is significant beyond the uncertainty. While this velocity difference is smaller than the spectral element scale by a factor of $\sim 5$, such a sub-pixel scale difference is commonly reliable for the high SNR lines (e.g., \citealt{TeagueForemanMackey18}). In mN, further resolved motion can be seen: redshifted (blueshifted) in the southwest (northeast) part. This may suggest a rotation or smaller size substructures in mN. 

In the right panel of Figure~\ref{fig:mommaps}, there is a velocity dispersion peak at the boundary of mN and sN1. In the velocity and velocity dispersion maps of Figure~\ref{fig:mommaps}, NIRCam F115W surface brightness is superposed by the contours. The F115W band samples the rest-frame $1500$\,\AA\ light and is sensitive to the 100\,Myr scale star formation. The velocity dispersion peak between mN and sN1 is interesting where no corresponding star formation nor line flux peak is observed (Section~\ref{sec:MultiPhot}).

We now investigate whether the kinematics of \tName\ are consistent with a merger or a rotation disk. We measured an observed velocity difference ($\Delta v_{\rm obs}$) and intrinsic velocity dispersion ($\sigma_0$) following an approach applied to IFS data from the KMOS$^{\rm 3D}$ survey of $0.7 \lesssim z \lesssim 2.7$ galaxies \citep{Wisnioski+15,Wisnioski+19,Lang+17,ForsterSchreiber+19}. The $\Delta v_{\rm obs}$ corresponds to the rotation velocity if the system is an edge-on rotating disk. The $\sigma_0$ is a representative velocity dispersion per pixel taking account of the beam smearing effect \citep{ForsterSchreiber+09,ForsterSchreiber+14,NewmanGenzel+13}, which is practically measured at the galaxy outskirt and gets smaller than a simple average of the velocity dispersion map. Details are described in Appendix~\ref{ap:kinematics}. The measured $\Delta v_{\rm obs}$ and $\sigma_0$ are $34$\,km\,s$^{-1}$ and $\sim 30$\,km\,s$^{-1}$, respectively. 

We applied criteria to judge a rotating disk proposed by \citet{Wisnioski+15}: 1) the given galaxy exhibits a continuous velocity gradient along a single axis, 2) rotationally-supported $v_{\rm rot} / \sigma_0 > 1$, 3) the position of the steepest velocity gradient is coincident with the velocity dispersion peak, 4) the photometric and kinematic major axes are in agreement, and 5) the position of the steepest velocity gradient is coincident with the centroid of the continuum. 
The rotation velocity is expressed as $v_{\rm rot} = \Delta v_{\rm obs} / \sin(i)$, where $i$ is an inclination of the galaxy. The measured $v_{\rm rot} / \sigma_0 = 1.1 / \sin(i)$ satisfies the second criterion, regardless of the inclination $i$. The position of the steepest velocity gradient is the midpoint between mN and sN1 (the kinematic center defined in Appendix~\ref{ap:kinematics}). The kinematic major axis is defined as the direction of the largest observed velocity difference that passes through sE, mN, and sN1 (Appendix~\ref{ap:kinematics}). The photometric axis (the stellar axis in Appendix~\ref{ap:kinematics}) aligns the F115W surface brightness or stellar distribution penetrating the two continuum centroids at mN and mS. \tName\ satisfies only the criteria 2) and 3). We note that the frequently used second criterion involves uncertainties in not only the measurements but also the threshold. \tName\ does not satisfy the modified version of the second criterion, $v_{\rm rot} / \sigma_0 > \sqrt{3.36}$ \citep{ForsterSchreiber+18,Wisnioski+19}, if the inclination $i > 37$\,degrees. Also, some authors use $\Delta v_{\rm obs} / \sigma_{\rm tot} > 0.4$ with $\sigma_{\rm tot}$ measured from spatially integrated spectrum (e.g., \citealt{Smit+18,Tokuoka+22}). Combining the multiple criteria is essentially important for robust kinematic classification, which is made possible by high-resolution IFS instruments such as JWST NIRSpec. 

In our NIRSpec IFS data, \tName\ is not classified as a purely rotation-dominated disk. The multiple clumps in the \OIII\,5008\AA\ emission can be interpreted as merger-related structures, either merging galaxies or clumps formed through tidal compression during a merger \citep[e.g.,][]{Nakazato+24}. Alternatively, they can be star-forming clumps induced by disk instability within a single galaxy. In the latter case, the underlying disk is hard to be observed via the \OIII\,5008\AA\ emission because local star-formation in the clumps disturbs warm ionized gas. It is observationally difficult to further distinguish the physical origins of the multiple clumps. We will revisit this issue with simulations and previous ALMA works in Section~\ref{sec:disc_view}.

\subsection{Spatially Resolved Line Diagnostics} \label{sec:lineRatio}

\begin{figure*}[htbp]
    \centering
    \includegraphics[width=1.0\linewidth, angle=0]{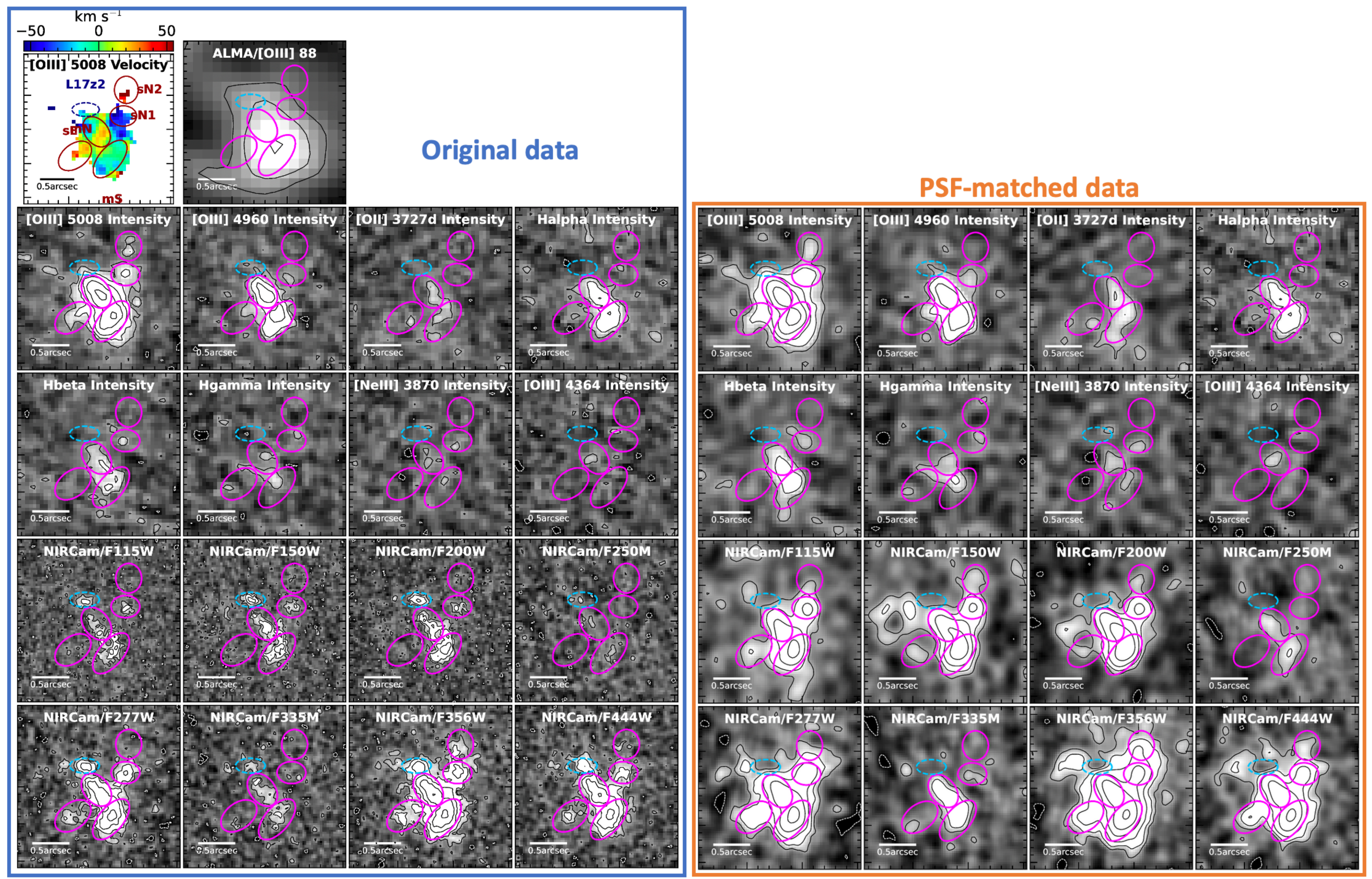}
    \caption{Multiple clumps identified in the NIRSpec line integrated intensity maps and NIRCam images. We also show the \OIII\,5008\AA\ velocity and ALMA \OIII\,88$\mu$m integrated intensity maps at the top left panels. Not only the original data but also the PSF-matched data are shown in the left- and right-side panels, respectively. Each panel has $2\farcs2 \times 2\farcs2$ in size. In the all panels, the photometric apertures for the multiple clumps are shown by magenta ellipses. Contour levels are $-2$ (dashed), 2, 4, 8, 16, and $32 \times \sigma$ (solid). The low-$z$ contaminant, L17z2, is masked when generating the PSF-matched NIRCam images (see Section~\ref{sec:MultiPhotClump}). }
    \label{fig:stamps}
\end{figure*}

\begin{figure*}[]
    \centering
    \includegraphics[width=1.0\linewidth, angle=0]{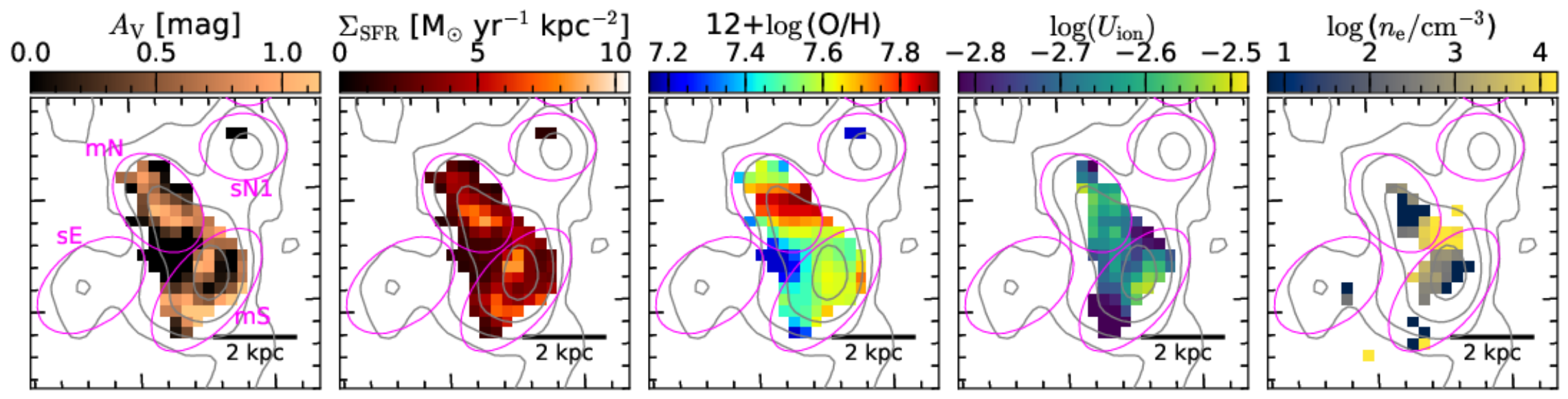}
    \caption{Maps of physical quantities: dust attenuation $A_{\rm V}$, 10\,Myr time-scale (H$\alpha$) star formation rate surface density $\Sigma_{\rm SFR}$, metallicity $12 + \log({\rm O/H})$, ionizing parameter $\log(U_{\rm ion})$, and electron density $\log(n_{\rm e}/{\rm cm}^{-3})$ from left to right. Pixel-to-pixel physical properties of $A_{\rm V}$, SFR, $12+\log({\rm O/H})$, $\log(U_{\rm ion})$, and $\log(n_{\rm e})$ are derived from Balmer decrement (H$\alpha$/H$\beta$), H$\alpha$, R3 index (\OIII\,5008\AA, H$\beta$), O32 index (\OIII5008\AA, \OII\ doublet), and \OII\,3727,3730\AA\ line ratio, respectively. Each panel covers 6.9 $\times$ 6.9 kpc at $z = 6.81$. The multiple clumps in \tName\ are shown by ellipses. Contours show the PSF-matched F115W surface brightness profile at 2, 4, 8, and 16 $\times \sigma$ levels. The F115W surface brightness is sensitive to $100$\,Myr time-scale star formation. }
    \label{fig:PhysParamMaps}
\end{figure*}

The NIRSpec line integrated intensity maps (Section~\ref{sec:LineSearch1}) are shown in the left panels of Figure~\ref{fig:stamps}. To investigate emission line flux ratios pixel-by-pixel, we homogenized PSFs of the line integrated intensity maps to that of the NIRSpec H$\alpha$ integrated intensity map with FWHM $=$ 0\farcs21 (PSF-matched data; see Appendix~\ref{ap:PSFmatch}). These PSF-matched line integrated intensity maps are shown in the right panels of Figure~\ref{fig:stamps}. From the PSF-matched images, we produced maps of physical quantities: dust attenuation ($A_{\rm V}$), 10\,Myr time-scale star formation rate surface density ($\Sigma_{\rm SFR}$), metallicity ($12 + \log({\rm O/H})$), and ionization parameter ($\log U_{\rm ion}$). We also produced a map of electron density ($\log(n_{\rm e}/{\rm cm^{-3}})$) using the NIRSpec IFU data cube before applying the PSF-matching, based on the \OII\,3729,3730\AA\ doublet. These maps are shown in Figure~\ref{fig:PhysParamMaps}. In every panel, we superposed the PSF-matched F115W surface brightness in contours which traces the rest-frame $1500$\,\AA\ continuum and then 100\,Myr time-scale SFR surface density. 

The dust attenuation $A_{\rm V}$ was measured from the Balmer decrement H$\alpha/$H$\beta$ assuming the SMC dust extinction law \citep{Prevot+84,Gordon+03} and the intrinsic line flux ratio of 2.76\footnote{The intrinsic H$\alpha$/H$\beta = 2.76$ is the value under assumption of electron temperature of $20,000$\,K \citep{Luridiana+15,StoreyHummer95}. This temperature is obtained for \tName\ in the companion paper \citep{Usui+25}.}. We adopted the SMC law, as it is likely applicable to high-$z$ low-mass galaxies (\citealt{Reddy+15,Fudamoto+17,Fudamoto+20}; but see also \citealt{Scoville+15,McLure+18}) and is also supported in the case of \tName\ (Section~\ref{sec:IRXbeta}). We masked out pixels with SNR lower than 2 in the H$\alpha$ or H$\beta$ integrated intensity map. The $A_{\rm V}$ map in Figure~\ref{fig:PhysParamMaps} clearly shows spatial variation of the dust attenuation. From comparison with the F115W surface brightness (contours in Figure~\ref{fig:PhysParamMaps}), the dust attenuation is as high as $A_{\rm V} \sim 1$\,mag near the rest-frame UV emission (i.e., young stellar populations) peaks of mN and mS. The $A_{\rm V}$ peak values are much higher than $A_{\rm V}$ estimated from the H$\alpha$ and H$\beta$ fluxes spatially integrated within the clumps: $A_{\rm V} = 0.28^{+0.38}_{-0.28}$\,mag and $0.22^{+0.35}_{-0.22}$\,mag for mN and mS, respectively. In the case that dust is distributed inhomogeneously, dust extinction measurements from the spatially integrated H$\alpha$/H$\beta$ ratio tend to be biased toward low-extinction regions, underestimating the true amount of dust attenuation \citep{Kreckel+13,Nelson+16}.

The $\Sigma_{\rm SFR}$ was measured from the H$\alpha$ flux using a conversion formula used in \citet{Shapley+23}, $\mathrm{SFR} = (2.14 \times 10^{42}) \times L({\rm H}\alpha)$, which is tuned for low metallicity ($Z = 0.27\,Z_{\odot}$) galaxies under Chabrier IMF. The H$\alpha$ emission traces 10\,Myr time-scale star-formation. We here corrected for the dust attenuation using the $A_{\rm V}$ map. The SFR map in Figure~\ref{fig:PhysParamMaps} shows the similar trend as the $A_{\rm V}$ map: the $\Sigma_{\rm SFR}$ peaks almost align peaks of young stars (F115W) in mN and mS. The peak and averaged $\Sigma_{\rm SFR}$ are $\sim 9$ and $\sim 4$\,M$_{\odot}$\,yr$^{-1}$\,kpc$^{-2}$, respectively, both of which are broadly consistent with $\Sigma_{\rm SFR}$ of $5 < z < 14$ galaxies and well above the $z < 3$ measurements (\citealt{Morishita+24a} and references therein). 

The metallicity $12 + \log({{\rm O/H}})$ was estimated from the R3 index following \citet{Hirschmann+23}, where R3 $=$ \OIII\,5008\AA$/$H$\beta$. All the line fluxes were corrected for dust attenuation using the $A_{\rm V}$ map. The metallicity map of Figure~\ref{fig:PhysParamMaps} shows variation of $12 + \log({\rm O/H})$ from $7.2$ to $8$. These values are comparable to metallicity measurements of $3 < z < 9.1$ galaxies \citep{Morishita+24b}. In mN the metallicity is $12 + \log({\rm O/H}) \sim 7.8$, higher than mS and other regions with $\lesssim 7.6$. The metallicty difference between mN and the other regions is significant beyond the uncertainty that was calculated by propagating the line flux errors in each pixel. These internal metallicity variations of $\sim 0.6$\,dex are similar as those found in other high-$z$ groups of merging galaxies \citep{Arribas+24,Venturi+24}. The mN and mS clumps may have originally been distinct galaxies with  different evolutionary paths (and then different metal enrichment histories).

The ionization parameter is defined as the ratio of ionizing photon flux to hydrogen atom density \citep{KewleyDopita02} and often expressed as a dimensionless form, $U_{\rm ion}$, by dividing by the speed of light. We used an analytic formulae of \citet{NakajimaOuchi14} using $12 + \log({\rm O/H})$ and O32 index, where O32 $=$ \OIII\,4960,5008\AA$/$\OII\,3727,3730\AA. All the line fluxes were corrected for dust attenuation using the $A_{\rm V}$ map. 
In the $\log U_{\rm ion}$ map in Figure~\ref{fig:PhysParamMaps}, while the map spatial coverage is limited by the \OII\,3727,3730 SNR, we can see spatial variation from $\log U_{\rm ion} = -2.9$ to $-2.5$. The ionization parameters at any location in \tName\ are lower than the median value of $z = 7$ -- $9$ galaxies in the CEERS survey, $\log U_{\rm ion} = -2.1$ \citep{Tang+23}, and the typical value of $z = 5.5$ -- $9.5$ galaxies in the JADES survey, $\log U_{\rm ion} = -1.5$ \citep{Cameron+23}

We derived the map of electron density, log($n_{\rm e}$/cm$^{-3}$), based on the line ratio of the \OII\,3729,3730 doublet \citep{AGNAGN}. For each spaxel in the NIRSpec IFS cube, we extracted the spectrum around the \OII\ doublet and performed a double Gaussian fitting with MC iterations, following the same procedure as in Section~\ref{sec:LineSearch1}. From the best-fit line fluxes, we calculated the line ratio and converted it to $n_{\rm e}$. The resulting electron density map is shown in the rightmost panel of Figure~\ref{fig:PhysParamMaps}. Interestingly, the derived electron densities span more than three orders of magnitude, ranging from $n_{\rm e} \sim 10$ to $10^4$\,cm$^{-3}$. This large variation in $n_{\rm e}$ originates from the measured \OII\ line ratios spanning the full theoretical range, from $\sim 0.3$ to $\sim 1.5$ \citep{AGNAGN}. The density variations are significant beyond the uncertainties. Notably, the highest electron densities are found in the region between mN and mS (and possibly sN1), which may suggest that the enhanced densities are related to an ongoing merger between these clumps \citep[c.f.,][]{Cox+06}.

The physical quantities investigated in this section roughly correlate with the spatial distribution of young stars traced by the F115W surfae brightness. At the F115W peaks in mN and mS, star formation is more active, dust attenuation is higher, and ionization parameter is higher than at the outskirt regions. The notable exception comes from the metallicity map where mN has higher values than mS by $\sim 0.2$\,dex. Mapping of the physical quantities in sub-clumps sE, sN1, and sN2 will help the correct understanding, which requires deeper IFS observations in future.

\section{Multi-Wavelength Analysis of the Clumps} \label{sec:multiwaveana}

\subsection{Multiband Photometry} \label{sec:MultiPhot}

\subsubsection{Individual Clumps} \label{sec:MultiPhotClump}

Figure~\ref{fig:stamps} shows the NIRCam images and ALMA \OIII\,88$\mu$m integrated intensity map as well as the NIRSpec line integrated intensity maps. The five clumps, mN, mS, sE, sN1, and sN2 (magenta ellipses in Figure~\ref{fig:stamps}), are well resolved in many bands and line maps except for the ALMA \OIII\,88$\mu$m map that has a coarser spatial resolution. While mN and mS are further divided into smaller structures in the NIRCam F115W image, we investigate the five clumps defined in Section~\ref{sec:kinematics} in the subsequent multi-wavelength analyses. 

Before focusing on the five clumps, we mention that there is a $z \sim 2$ interloper \citep{Laporte+17,Smit+15} lying at northeast of mN (cyan dashed ellipse in Figure~\ref{fig:stamps})\footnote{We checked Subaru public optical images from the Subaru strategic program with HSC (HSC-SSP; \citealt{Aihara+19,Furusawa+18,Kawanomoto+18,Komiyama+18,Miyazaki+18}), and found that L17z2 is detected in the $g$, $r$, $i$, and $z$ bands with significance of 2 -- 3\,$\sigma$. This confirms that L17z2 is a lower-$z$ interloper.}. 
We hereafter refer to this object as L17z2. In the following analyses, we masked L17z2 in the NIRCam images. We did not mask it in the NIRSpec data because the low-$z$ interloper should not emit line emission at the specific wavelengths. Actually L17z2 cannot be seen in the NIRSpec line integrated intensity maps (Figure~\ref{fig:stamps}). 

For the multiband SED analyses, we used the PSF-matched NIRSpec and NIRCam data (Section~\ref{sec:kinematics}; Appendix~\ref{ap:PSFmatch}).
From the PSF-matched NIRSpec data cube, we extracted one-dimensional spectra of the five clumps whose apertures are defined as the solid ellipses in Figure~\ref{fig:mommaps} and \ref{fig:stamps}. In the same manner as in Section~\ref{sec:LineSearch1}, we measured the \OII\,3727\AA, \OII\,3730\AA, \NeIII\,3870\AA, H$\gamma$, H$\beta$, \OIII\,4960\AA, \OIII\,5008\AA, and H$\alpha$ line fluxes and the uncertainties of the five clumps. 
We put lower limit on the flux uncertainty as 10\,\% of the flux to securely take account of the absolute flux uncertainty \citep{Boker+23}\footnote{See also ``JWST Absolute Flux Calibration'' in JWST User Documentation: \url{https://jwst-docs.stsci.edu/jwst-calibration-status/jwst-absolute-flux-calibration}}. 
We determined detection or nondetection for each line of each clump according to the three criteria introduced in Section~\ref{sec:LineSearch1}. All the line fluxes are shown in Table~\ref{tb:fluxAll}. We put $3\,\sigma$ upper limits for nondetections. In Table~\ref{tb:fluxAll}, we also show the redshifts from \OIII\,5008\AA\ and the radial velocity offsets with respect to $z = 6.80731$ that is the fiducial estimate in the central region (Section~\ref{sec:LineSearch1}).

For the NIRCam flux density of the five clumps, we performed aperture photometry using \texttt{photutils} \citep{Bradley+23} on the PSF-matched images (the right panels of Figure~\ref{fig:stamps}). We subtracted from the aperture flux local sky background which was measured using a circular annulus aperture with inner/outer radius of $1''$/$2''$. The photometric uncertainties were estimated from random aperture photometry described in Section~\ref{sec:NIRCAMobs}. In addition, we put lower limit of 5\,\% on the photometric uncertainty to securely take account of the absolute flux uncertainty \citep{Rigby+23}. All bands flux densities are shown in Table~\ref{tb:fluxAll}.
We adopted $3\,\sigma$ upper limits for flux densities below the $3\,\sigma$ detection threshold.

While the \OIII\,88$\mu$m line is significantly detected in ALMA Band\,8, the relatively coarse resolution does not allow us to spatially resolve mN, mS, sE, sN1, and sN2 clumps (Figure~\ref{fig:stamps}). We then measured only the total \OIII\,88$\mu$m line flux (Section~\ref{sec:MultiPhotTot}).
Dust continuum emission is not detected in Band\,5, 6, nor 8. 
Sizes of the five clumps are smaller than the beam sizes of the ALMA data. Then, we set three times the noise rms per beam in the ALMA continuum images (Section~\ref{sec:ALMAdata}) for the flux density upper limit of each clump. 
We further applied the correction for Cosmic Microwave Background (CMB) effect \citep{daCunha+13,Ota+14} assuming dust temperature $T_{\rm d} = 50$\,K and emissivity index $\beta_{\rm d} = 2.0$ \citep[e.g.,][]{Sommovigo+22a,Sommovigo+22b}. The CMB correction factor is 1.13, 1.08, and 1.01 in Band\,5, 6, and 8, respectively. All the ALMA continuum flux densities are summarized in Table~\ref{tb:fluxAll}.

\begin{table*}[]
\begin{center}
\caption{Observed measurements in NIRSpec IFU, NIRCam, and ALMA data. The $3\,\sigma$ upper limits are given for nondetections.} \label{tb:fluxAll}
\begin{tabular}{ccccccc}
\hline
\hline
  & mN &  mS &  sE &  sN1 &  sN2 & Total \\ 
\hline
Redshift\tablenotemark{\rm a}  &  6.80745 & 6.80714 & 6.80758 & 6.80648 & 6.80903 & 6.80726 \\ 
   &  $\pm 0.00002$ & $\pm 0.00002$ & $\pm 0.00013$ & $\pm 0.00009$ & $\pm 0.00019$ &  $\pm 0.00002$ \\ 
Velocity\tablenotemark{\rm b}  &  \multirow{2}{*}{5.3 $\pm$ 0.7} & \multirow{2}{*}{-6.4 $\pm$ 0.8} & \multirow{2}{*}{10.3 $\pm$ 4.9} & \multirow{2}{*}{-32.1 $\pm$ 3.4} & \multirow{2}{*}{66.0 $\pm$ 7.4} & \multirow{2}{*}{-2.1 $\pm$ 0.9} \\ 
 (km s$^{-1}$)  &  &  &  &  &  &   \\ 
FWHM\tablenotemark{\rm c}  &  \multirow{2}{*}{134.8 $\pm$ 2.0} & \multirow{2}{*}{149.0 $\pm$ 2.2} & \multirow{2}{*}{116.4 $\pm$ 13.8} & \multirow{2}{*}{70.3 $\pm$ 8.0} & \multirow{2}{*}{43.9 $\pm$ 10.8} & \multirow{2}{*}{144.0 $\pm$ 2.6} \\ 
 (km s$^{-1}$)  &  &  &  &  &  &   \\ 
$F_{\rm [OII] 3727}$  &  \multirow{2}{*}{1.33 $\pm$ 0.3} & \multirow{2}{*}{1.89 $\pm$ 0.25} & \multirow{2}{*}{0.51 $\pm$ 0.19} & \multirow{2}{*}{$<$ 0.44} & \multirow{2}{*}{$<$ 0.3} & \multirow{2}{*}{7.12 $\pm$ 0.87} \\ 
 (10$^{-18}$ erg s$^{-1}$ cm$^{-2}$)  &   &  &  &   &   &   \\ 
$F_{\rm [OII] 3730}$  &  \multirow{2}{*}{1.83 $\pm$ 0.31} & \multirow{2}{*}{2.22 $\pm$ 0.39} & \multirow{2}{*}{0.63 $\pm$ 0.22} & \multirow{2}{*}{$<$ 0.46} & \multirow{2}{*}{$<$ 0.46} & \multirow{2}{*}{7.6 $\pm$ 1.17} \\ 
 (10$^{-18}$ erg s$^{-1}$ cm$^{-2}$)  &   &  &  &   &   &   \\ 
$F_{\rm [NeIII] 3870}$  &  \multirow{2}{*}{1.22 $\pm$ 0.2} & \multirow{2}{*}{1.17 $\pm$ 0.18} & \multirow{2}{*}{$<$ 0.37} & \multirow{2}{*}{$<$ 1.48} & \multirow{2}{*}{$<$ 0.53} & \multirow{2}{*}{4.39 $\pm$ 0.57} \\ 
 (10$^{-18}$ erg s$^{-1}$ cm$^{-2}$)  &   &  &   &   &   &   \\ 
$F_{\rm H\gamma}$  &  \multirow{2}{*}{1.08 $\pm$ 0.49} & \multirow{2}{*}{1.11 $\pm$ 0.23} & \multirow{2}{*}{$<$ 1.55} & \multirow{2}{*}{$<$ 0.37} & \multirow{2}{*}{$<$ 0.29} & \multirow{2}{*}{2.53 $\pm$ 0.97} \\ 
 (10$^{-18}$ erg s$^{-1}$ cm$^{-2}$)  &   &  &   &   &   &   \\ 
$F_{\rm [OIII] 4364}$  &  \multirow{2}{*}{1.01 $\pm$ 0.33} & \multirow{2}{*}{$<$ 0.53} & \multirow{2}{*}{$<$ 0.32} & \multirow{2}{*}{$<$ 0.26} & \multirow{2}{*}{$<$ 0.05} & \multirow{2}{*}{1.83 $\pm$ 0.5} \\ 
 (10$^{-18}$ erg s$^{-1}$ cm$^{-2}$)  &   &   &   &   &   &   \\ 
$F_{\rm H\beta}$  &  \multirow{2}{*}{2.02 $\pm$ 0.2} & \multirow{2}{*}{2.32 $\pm$ 0.23} & \multirow{2}{*}{$<$ 0.47} & \multirow{2}{*}{0.34 $\pm$ 0.39} & \multirow{2}{*}{$<$ 0.24} & \multirow{2}{*}{7.98 $\pm$ 0.79} \\ 
 (10$^{-18}$ erg s$^{-1}$ cm$^{-2}$)  &   &  &   &  &  &   \\ 
$F_{\rm [OIII] 4960}$  &  \multirow{2}{*}{5.22 $\pm$ 0.52} & \multirow{2}{*}{5.36 $\pm$ 0.54} & \multirow{2}{*}{0.76 $\pm$ 0.11} & \multirow{2}{*}{0.75 $\pm$ 0.12} & \multirow{2}{*}{$<$ 0.25} & \multirow{2}{*}{17.4 $\pm$ 1.8} \\ 
 (10$^{-18}$ erg s$^{-1}$ cm$^{-2}$)  &   &  &  &  &   &   \\ 
$F_{\rm [OIII] 5008}$  &  \multirow{2}{*}{15.37 $\pm$ 1.54} & \multirow{2}{*}{16.46 $\pm$ 1.65} & \multirow{2}{*}{2.31 $\pm$ 0.23} & \multirow{2}{*}{1.78 $\pm$ 0.18} & \multirow{2}{*}{0.85 $\pm$ 0.09} & \multirow{2}{*}{52.3 $\pm$ 5.43} \\ 
 (10$^{-18}$ erg s$^{-1}$ cm$^{-2}$)  &   &  &  &  &  &   \\ 
$F_{\rm H\alpha}$  &  \multirow{2}{*}{6.19 $\pm$ 0.62} & \multirow{2}{*}{6.98 $\pm$ 0.7} & \multirow{2}{*}{1.0 $\pm$ 0.25} & \multirow{2}{*}{0.82 $\pm$ 0.38} & \multirow{2}{*}{$<$ 0.53} & \multirow{2}{*}{23.9 $\pm$ 2.36} \\ 
 (10$^{-18}$ erg s$^{-1}$ cm$^{-2}$)  &   &  &  &  &   &   \\ 
$f_{\nu,{\rm F115W}}$ ($\mu$Jy)  &  0.057 $\pm$ 0.005 & 0.103 $\pm$ 0.006 & 0.021 $\pm$ 0.006 & 0.032 $\pm$ 0.004 & $<$ 0.014 & 0.351 $\pm$ 0.046 \\ 
 $f_{\nu,{\rm F150W}}$ ($\mu$Jy)  &  0.071 $\pm$ 0.005 & 0.12 $\pm$ 0.006 & $<$ 0.017 & 0.028 $\pm$ 0.004 & $<$ 0.013 & 0.354 $\pm$ 0.046 \\ 
 $f_{\nu,{\rm F200W}}$ ($\mu$Jy)  &  0.077 $\pm$ 0.004 & 0.12 $\pm$ 0.006 & 0.014 $\pm$ 0.005 & 0.028 $\pm$ 0.003 & $<$ 0.011 & 0.369 $\pm$ 0.04 \\ 
 $f_{\nu,{\rm F250M}}$ ($\mu$Jy)  &  0.081 $\pm$ 0.021 & 0.111 $\pm$ 0.026 & $<$ 0.07 & $<$ 0.049 & $<$ 0.056 & 0.458 $\pm$ 0.167 \\ 
 $f_{\nu,{\rm F277W}}$ ($\mu$Jy)  &  0.089 $\pm$ 0.004 & 0.131 $\pm$ 0.007 & 0.029 $\pm$ 0.003 & 0.036 $\pm$ 0.002 & 0.01 $\pm$ 0.002 & 0.479 $\pm$ 0.025 \\ 
 $f_{\nu,{\rm F335M}}$ ($\mu$Jy)  &  0.103 $\pm$ 0.013 & 0.11 $\pm$ 0.016 & $<$ 0.042 & $<$ 0.03 & $<$ 0.034 & 0.472 $\pm$ 0.096 \\ 
 $f_{\nu,{\rm F356W}}$ ($\mu$Jy)  &  0.168 $\pm$ 0.008 & 0.226 $\pm$ 0.011 & 0.043 $\pm$ 0.003 & 0.05 $\pm$ 0.003 & 0.024 $\pm$ 0.002 & 0.836 $\pm$ 0.042 \\ 
 $f_{\nu,{\rm F444W}}$ ($\mu$Jy)  &  0.128 $\pm$ 0.006 & 0.175 $\pm$ 0.009 & 0.036 $\pm$ 0.004 & 0.04 $\pm$ 0.003 & 0.014 $\pm$ 0.003 & 0.619 $\pm$ 0.031 \\
 $F_{\rm [OIII] 88}$  &  \multirow{2}{*}{---} & \multirow{2}{*}{---} & \multirow{2}{*}{---} & \multirow{2}{*}{---} & \multirow{2}{*}{---} & \multirow{2}{*}{15.37 $\pm$ 3.33} \\ 
 (10$^{-18}$ erg s$^{-1}$ cm$^{-2}$)  &   &  &  &  &  &  \\ 
$f_{\nu,{\rm Band5}}$ ($\mu$Jy)  &  $<$ 22.0 & $<$ 22.0 & $<$ 22.0 & $<$ 22.0 & $<$ 22.0 & $<$ 34.5 \\ 
$f_{\nu,{\rm Band6}}$ ($\mu$Jy)  &  $<$ 19.4 & $<$ 19.4 & $<$ 19.4 & $<$ 19.4 & $<$ 19.4 & $<$ 48.6 \\ 
$f_{\nu,{\rm Band8}}$ ($\mu$Jy)  &  $<$ 213.1 & $<$ 213.1 & $<$ 213.1 & $<$ 213.1 & $<$ 213.1 & $<$ 344.8 \\ 
\hline
\end{tabular}
\end{center}
\tablenotetext{\rm a}{Redshifts estimated by the Gaussian profile fitting to the NIRSpec \OIII\,5008\AA\ line.}
\tablenotetext{\rm b}{Radial velocity offset} relative to $z = 6.80731$, which is estimated from the redshifts of the \OIII\,5008\AA\ lines.
\tablenotetext{\rm c}{Line FWHMs estimated by the Gaussian profile fitting to the NIRSpec \OIII\,5008\AA\ line. They are corrected for the instrumental broadening \citep{Jakobsen+22}. }
\tablenotetext{\rm c}{Sum of the \OII\,3727\AA\ and \OII\,3730\AA\ fluxes. }
\tablenotetext{\rm d}{Upper limits on the ALMA continuum flux density are evaluated from the noise rms levels per beam for the individual clumps. For the total photometry, we corrected the noise rms per beam for areal difference between the beam and \OIII\,5008\AA\ emitting region (see text). The CMB effects \citep{daCunha+13,Ota+14} are also corrected. }
\end{table*}

\subsubsection{Total Photometry}\label{sec:MultiPhotTot}

To obtain total NIRSpec line fluxes of \tName, we extract a spectrum from the PSF-matched data cube with a $2\,\sigma$ isophotal aperture in the \OIII\,5008\AA\ integrated intensity map (the outermost contour in the PSF-matched \OIII\,5008\AA\ panel of Figure~\ref{fig:stamps}). Then, we measured the emission line fluxes in the same manner as in Section~\ref{sec:LineSearch1}. Although the isophotal aperture is expected to maximize the SNR for total flux (e.g., \citealt{Welch+24}), there may still be flux loss outside the aperture. We performed the curve-of-growth analysis (e.g., \citealt{Drake+17}) for the \OIII\,5008\AA\ flux, and found a $10$\,\% loss in the $2\,\sigma$ isophotal flux. We applied the aperture correction to the $2\,\sigma$ isophotal fluxes of the all lines assuming the same spatial distribution of ionized gas emitting the nebular lines. The resultant total line fluxes are shown in Table~\ref{tb:fluxAll}.

Because continuum morphology traced by the NIRCam bands is not necessarily similar to the nebular gas morphology, we estimated the total flux densities from convergence values of the curve-of-growths in the every individual NIRCam bands. They are shown in Table~\ref{tb:fluxAll}. Our NIRCam flux densities are consistent within the uncertainties with previous measurements using the HST and ground-based instruments \citep{Smit+15,Laporte+17,Nayyeri+17,Weaver+22} as well as the recent work using the PRIMER NIRCam data \citep{Harikane+25a}.

In ALMA data, we measured the total \OIII\,88$\mu$m line flux by two-dimensional Gaussian fitting using \texttt{CASA imfit} to the line integrated intensity map. The resultant line flux is $1.06 \pm 0.23$\,Jy\,km\,s$^{-1}$. \tName\ is spatially resolved in \OIII\,88$\mu$m and its beam-deconvolved size is $0\farcs88 \times 0\farcs45$. For the dust continuum, we estimated a conservative upper limit for the total flux density in each band by correcting the noise rms per beam (Section~\ref{sec:ALMAdata}) for the difference between the beam and object sizes \citep[e.g.,][]{Hashimoto+23b}. We here assumed that the size of the dust emission region equals to that of the \OIII\,88$\mu$m emission. We applied the CMB correction in the same manner as for the five clumps (Section~\ref{sec:MultiPhotClump}). The $3\,\sigma$ upper limits on the dust continuum flux densities are shown in Table~\ref{tb:fluxAll}.

\subsection{IRX--$\beta$ Relation} \label{sec:IRXbeta}

\begin{figure}
    \centering
    \includegraphics[width=1.0\linewidth, angle=0]{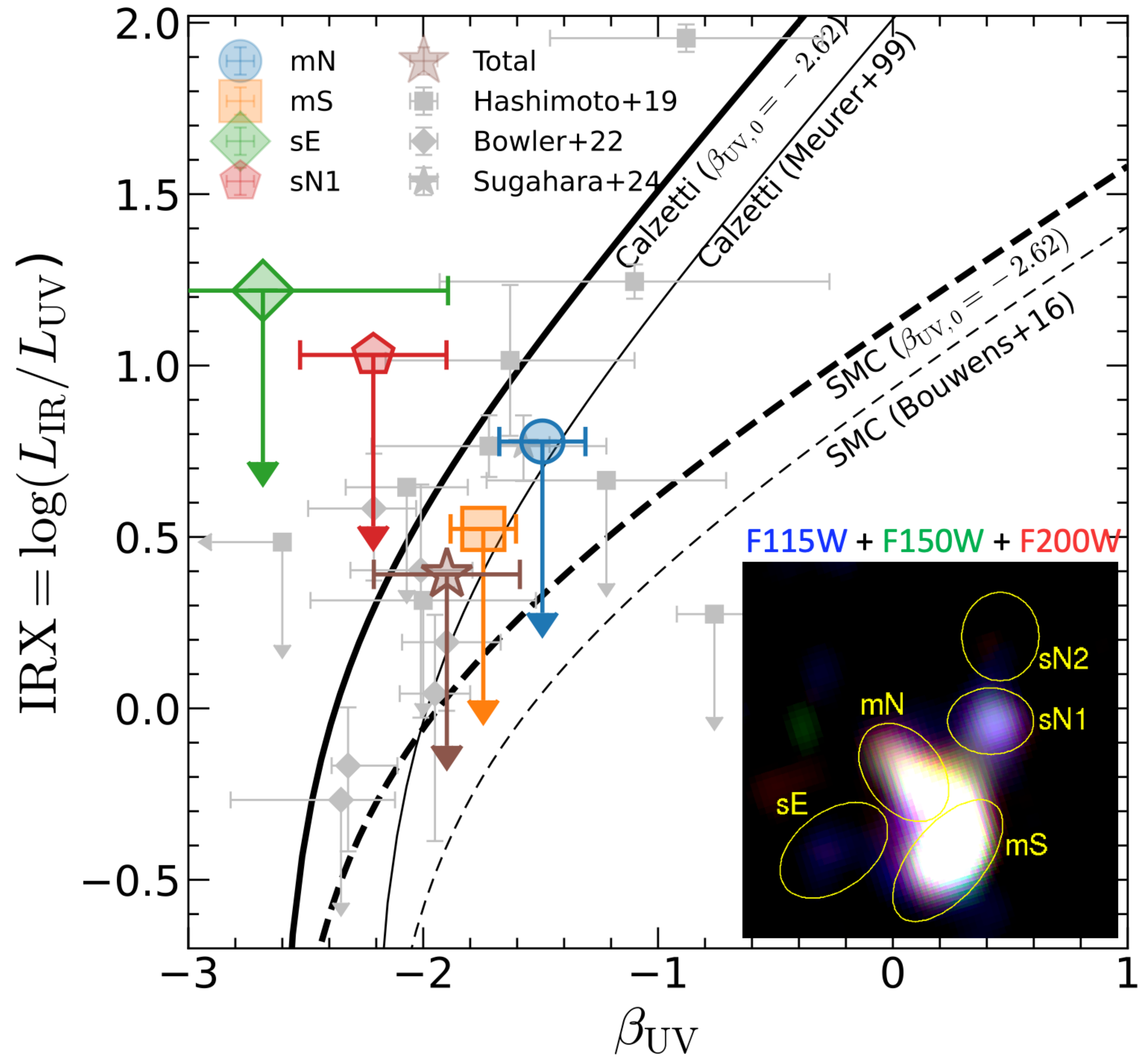}
    \caption{IRX--$\beta_{\rm UV}$ diagram. The mN, mS, sE, sN1 clumps, and total system are shown by the circle, square, pentagon, diamond and star symbols, respectively. Their IRX measurements are $3\,\sigma$ upper limits as represented by the arrows. The grey square, diamond, and star symbols show $6.5 < z < 9.1$ galaxies taken from \citet{Hashimoto+19a}, \citet{Bowler+22}, and \citet{Sugahara+25}, respectively. All the IRX values from the literature are corrected for $T_{\rm d} = 50$\,K and $\beta_{\rm d} = 2.0$. Thin solid and dashed lines correspond to the canonical relations assuming the Calzetti and SMC dust attenuarion curves \citep{Meurer+99,Bouwens+16c}, respectively. Thick lines are the same but modified with $\beta_{\rm UV,0} = -2.62$ (\citealt{Reddy+18}; see the text). The inset panel shows a F115W $+$ F150W $+$ F200W pseudo color image with the apertures for the multiple clumps. }
    \label{fig:IRX-beta}
\end{figure}

We investigate the relation between the infrared excess (IRX $= \log(L_{\rm IR} / L_{\rm UV})$, where $L_{\rm IR}$ and $L_{\rm UV}$ represent the UV and infrared luminosities, respectively) and $\beta_{\rm UV}$ to discuss whether the observed red $\beta_{\rm UV}$ and nondetection of the dust continuum are naturally explained with commonly used dust attenuation curves. 

The UV slope of \tName\ was measured by fitting the NIRCam F115W, F150W, and F200W flux densities (Section~\ref{sec:MultiPhot}) with an exponential function: $f_\lambda \propto \lambda^{\beta_{\rm UV}}$. These NIRCam bands sample the rest-frame wavelengths from 1300 to 2850\,\AA\ that is almost identical to the wavelength definition of \citet{Calzetti+01}. The UV luminosity is defined as $L_{\rm UV} = \lambda L_{\lambda}$ at $\lambda = 1600$\,\AA. We estimated $L_{\rm UV}$ from the F115W flux density. The IR luminosity $L_{\rm IR}$ was estimated by fitting the CMB-corrected ALMA Band\,5, 6, and 8 flux densities (Section~\ref{sec:MultiPhot}) with the modified black body assuming the dust temperature of $T_{\rm d} = 50$\,K and emissivity index of $\beta_{\rm d} = 2.0$. Because dust continuum is not detected in any bands, we estimated $3\,\sigma$ upper limits of $L_{\rm IR}$ for the multiple clumps. 

Figure~\ref{fig:IRX-beta} shows the IRX and $\beta_{\rm UV}$ of mN, mS, sE, sN1, and the total system (Total) by the circle, square, diamond, pentagon, and star symbols, respectively. For sN2, neither IRX nor $\beta_{\rm UV}$ can be estimated because of nondetection in F115W, F150W, F200W, and ALMA continuum bands. \tName, as a whole (Total), shows $\beta_{\rm UV} = -1.9 \pm 0.3$. Our measurement is bluer than that of the previous work ($\beta_{\rm UV} \sim -1.1$; \citealt{Smit+18}), which may be due to the difference of the instruments and filters used. The upper limit on the IRX of Total, $< 0.39$ ($3\,\sigma$), is significantly higher than that obtained in the previous work \citep{Smit+18}. This is due to our conservative upper limits on the ALMA continuum fluxes taking into account the aperture area (Section~\ref{sec:MultiPhot}). Looking at each clump, sN1 and sE have notably blue UV slope of $\beta_{\rm UV} = -2.2 \pm 0.3$ and $-2.7 \pm 0.8$, respectively, while mN and mS are as red as $\beta_{\rm UV} \sim -1.6$. This UV color difference can also be visually confirmed in the pseudo color image superposed in Figure~\ref{fig:IRX-beta}. This may suggest patchy stellar-and-dust geometry \citep{Faisst+17,Tamura+23}. Alternatively, sN1 and sE may be satellite galaxies with different dust properties compared to the main clumps. 

The IRX--$\beta_{\rm UV}$ relations are connected with attenuation curves because of the energy balance between absorption and reradiation by dust \citep{Meurer+99}. There are two types of frequently used ``canonical'' relations: one for local starbursts proposed by \citet{Meurer+99} that implicitly includes the Calzetti dust attenuation curve \citep{Calzetti+00,Bouwens+16c}, and another expected with the SMC attenuation curve \citep{Prevot+84,Gordon+03}. We hereafter refer them to the Calzetti and SMC relations for convenience. They are shown by the thin curves in Figure~\ref{fig:IRX-beta}, where we followed the calculation of \citet{Bouwens+16c}. We note that these canonical relations assume that the intrinsic UV slope without the dust attenuation is $\beta_{\rm UV,0} = -2.23$. High redshift galaxies tend to have bluer $\beta_{\rm UV,0}$ due to the young ages, low metallicities, and binary interactions \citep{Steidel+14,Steidel+16,Reddy+18}. We then modified the canonical relations so that they have $\beta_{\rm UV,0} = -2.62$ \citep{Reddy+18,Sugahara+25} which are shown by the bold curves in Figure~\ref{fig:IRX-beta}. 
In \tName, mN, mS, and Total prefer the SMC relation to the Calzetti relation. While the Calzetti relation can explain the measurements in \tName\ if $\beta_{\rm UV,0}$ is redder than $-2.2$, this possibility is rejected in our SED fitting analysis (Section~\ref{sec:SEDfit}). 

In Figure~\ref{fig:IRX-beta} we also plot spectroscopically confirmed galaxies at $6.5 < z < 9.1$ from \citet{Hashimoto+19a}, \citet{Bowler+22}, and \citet{Sugahara+25} where we corrected the literature data for the dust temperature difference. Most of the galaxies whose dust continuum emission is detected follow the Calzetti relation with $-2.6 < \beta_{\rm UV,0} < -2.2$ (see also \citealt{Hashimoto+19a,Bowler+22,Sugahara+25}). On the other hand, as seen in \tName, steeper attenuation curves including the SMC curve are favored in stacking analyses of UV-selected galaxies at $2 \lesssim z \lesssim 5$ (\citealt{Reddy+18,Fudamoto+20}; but see also the opposite conclusion in \citealt{Alvarez-Marquez+16,Alvarez-Marquez+19,Bowler+24}). There seems to be a diversity in the dust properties of high redshift galaxies. Difference between the galaxies on the Calzetti and SMC IRX--$\beta_{\rm UV}$ relations should be statistically investigated as a future work.

\subsection{SED Fitting} \label{sec:SEDfit}

\begin{figure*}
    \centering
    \includegraphics[width=0.9\linewidth, angle=0]{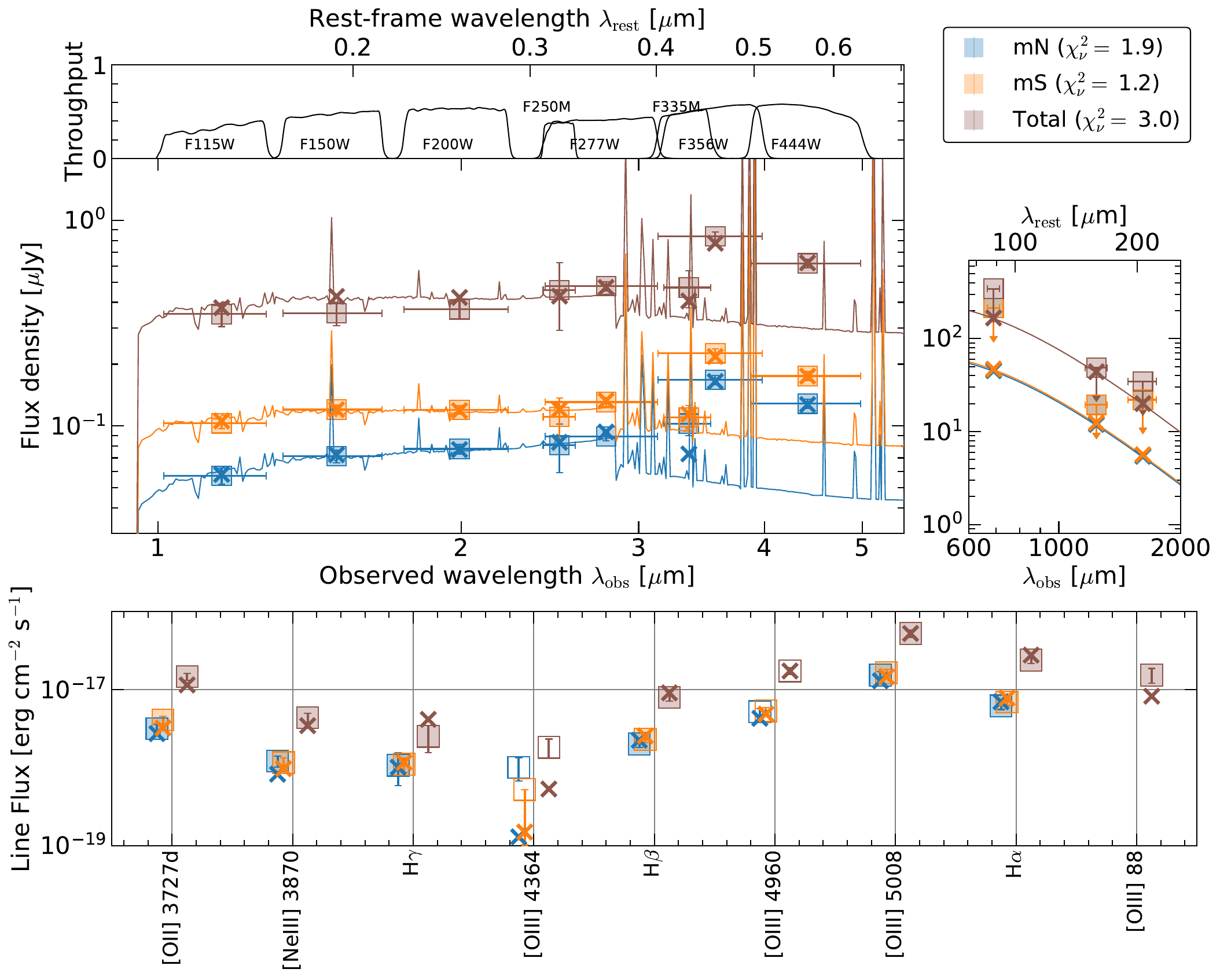}
    \caption{The observed SEDs and the best-fit models for mN, mS, and Total expressed by blue, orange, and brown symbols, respectively. In the top left panel, squares show the observed NIRCam flux density while the lines and crosses correspond to the best-fit model spectra and their filter-convolved flux density. The top right panel shows the same but for the ALMA dust continua. In the bottom panel, the squares and crosses correspond to the observed and best-fit model line fluxes, respectively. The label ``\OII\,3727d'' means sum of the \OII\,3727,3730\AA\ doublet flux. We also show the \OIII\,4364\AA\ and \OIII\,4960\AA\ lines that were actually not used in the SED fitting. ALMA \OIII\,88$\mu$m flux is shown only for the total photometry (Total) because of its coarse spatial resolution.}
    \label{fig:SEDres_SED_Mains}
\end{figure*}

\begin{figure*}
    \centering
    \includegraphics[width=1.0\linewidth, angle=0]{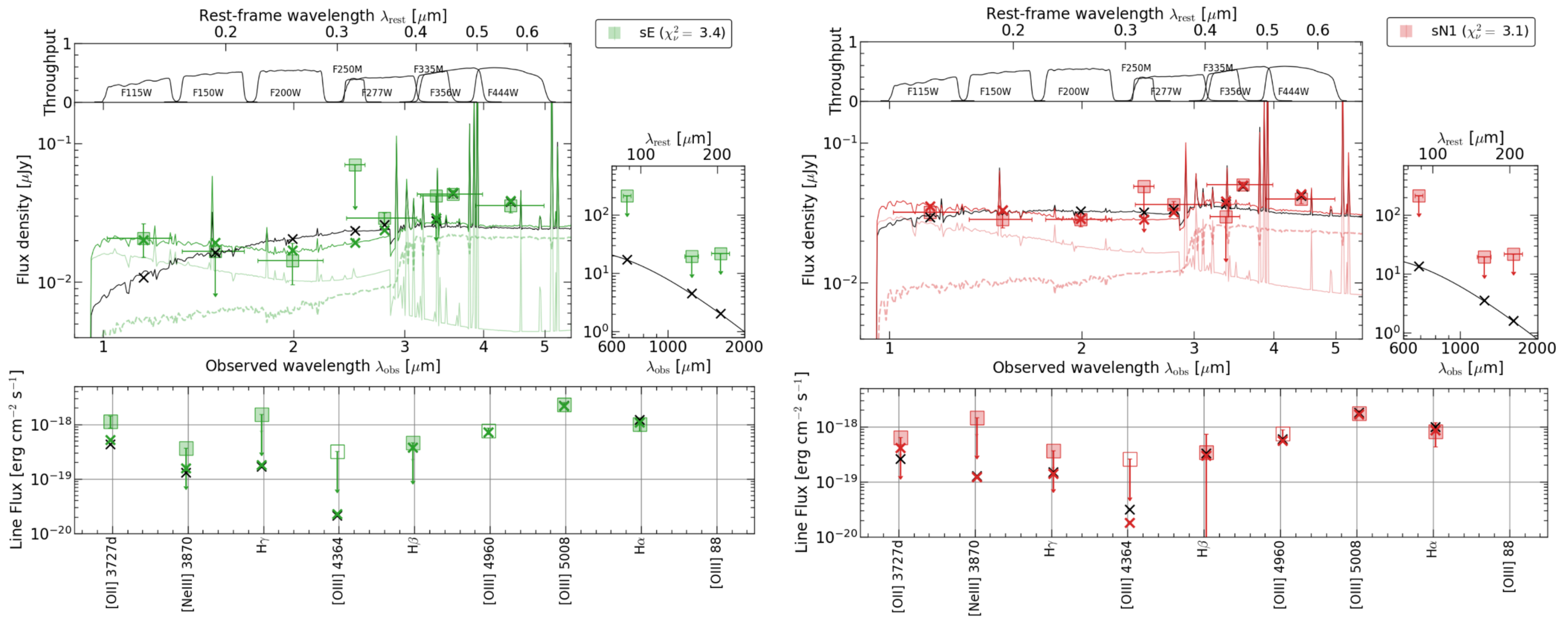}
    \caption{The observed SEDs and the best-fit models for sE (left panels) and sN1 (right panels). The two types of the fitting results are shown for each clump: single-component fitting (black solid spectrum and crosses) and old $+$ young two-components fitting (colored solid spectrum and crosses). In the latter fitting the old and young models are shown by thin dashed and solid spectra, respectively. }
    \label{fig:SEDres_SED_Subs}
\end{figure*}

We performed SED fitting analyses for the multiple clumps in \tName\ with the \texttt{PANHIT} code \citep{Mawatari+20,Mawatari+20b,Hashimoto+18a,Tamura+19}. The \texttt{PANHIT} combines the stellar, nebular, and dust emission SEDs by implementing the models of \citet{BruzualCharlot03}, \citet{Inoue+11b}, and the modified black body radiation, respectively. The \OIII\,88$\mu$m line flux is modeled with the empirical scaling relation from SFR \citep{Inoue+14b}. For the stellar models, we assumed the Chabrier IMF \citep{Chabrier03}. Fluxes at rest-frame wavelengths equal to or shorter than Ly$\alpha$ are attenuated by the
intergalactic medium (IGM), following the analytic model of \citet{Inoue+14a}. 

We simultaneously fit the NIRCam, ALMA continuum flux densities, NIRSpec, and ALMA emission line fluxes. We excluded the \OIII\,4960\AA\ and \OIII\,4364\AA\ line fluxes. The \OIII\,4960\AA\ flux degenerates with the \OIII\,5008\AA\ flux in the modeling because their ratio is fixed ($\simeq 3$; \citealt{StoreyZeippen00}). Combined with the \OIII\,5008 flux, the \OIII\,4364 flux is sensitive to the electron temperature $T_{\rm e}$ \citep{AGNAGN,Draine11}. The nebular emission model of \citet{Inoue+11b} assumes $T_{\rm e}$ as a function of metallicity, which limits the reproducibility of the temperature-sensitive line flux in galaxies that deviate from average physical condition. We performed detailed line diagnostics including the \OIII\,4364\AA\ line of \tName\ in our companion paper \citep{Usui+25}. The measured flux of the \OII\,3727,3730\AA\ doublet was summed so as to match the model that does not resolve the doublet \citep{Inoue+11b}. The \OIII\,88$\mu$m line flux was used only for the total photometry (Total) because it is not spatially resolved in ALMA Band\,8. On the other hand, the upper limits on the ALMA Band\,5, 6, and 8 flux densities were measured for each clump (Section~\ref{sec:MultiPhot}). 

In the SED fitting, we fixed the redshifts for the clumps (Section~\ref{sec:MultiPhot}). Star formation history (SFH) was assumed to be constant because it was not constrained in our trial with the delayed-$\tau$ model. Free parameters are dust attenuation $A_{\rm V}$, stellar age, metallicity $Z$, and escape fraction of ionizing photons $f_{\rm esc}$. Their fitting ranges are $0 \leq A_{\rm V} \leq 2$, 1\,Myr $<$ age $<$ the cosmic age at $z = 6.81$, $0.0001 \leq Z \leq 0.05$ ($0.005 Z_{\odot} \leq Z \leq 2.5 Z_{\odot}$), and $0 \leq f_{\rm esc} \leq 1$. We used the SMC dust attenuation curve that is preferred in the IRX--$\beta_{\rm UV}$ relation (Section~\ref{sec:IRXbeta}). Dust temperature and emissivity index in the modified black body are assumed to be $T_{\rm d} = 50$\,K and $\beta_{\rm d} = 2.0$. We assumed the same $A_{\rm V}$ for the nebular emission lines and stellar continua (e.g., \citealt{Erb+06c,Kashino+13}).

The best-fit model for the whole system (Total) is shown by the brown spectrum and crosses in Figure~\ref{fig:SEDres_SED_Mains}. We can see that the model reasonably reproduces the all observed measurements. The significant F356W excess, $\beta_{\rm UV} \sim -2$, and nondetection in ALMA dust continua are explained by combination of very young age ($8$\,Myr) and modest dust attenuation $A_{\rm V} \sim 0.2$. This consistency is broken if we assume lower dust temperature ($T_{\rm d} \leq 40$\,K) and less steep dust emissivity index ($\beta_{\rm d} \leq 1.5$). 

The best-fit models for the two bright main clumps mN and mS are shown by the blue and orange spectra and crosses in Figure~\ref{fig:SEDres_SED_Mains}. The observed SEDs are well reproduced and there are similar trends as in Total: the young age, modest dust attenuation, and high emission line equivalent widths (EWs). Their \OIII\,5008\AA\ EWs are as high as 1000 -- 2000\,\AA\ in the rest-frame, which are comparable to the highest values observed at $5 < z < 7$ \citep{Matthee+23}. 

The fitting results for the fainter sub-clumps sE and sN1 are shown in the left and right panels of Figure~\ref{fig:SEDres_SED_Subs}, respectively. Unfortunately, the best-fit solution for sN2 is unreliable due to the limited number of detections in its observed SED. For sE, we failed to obtain the reasonable solution even though this clump has five and four detections in NIRCam and NIRSpec which is enough to constrain the SED. The minimum $\chi^2$ model is shown by the black spectrum in Figure~\ref{fig:SEDres_SED_Subs}. The observed SED of sE is very different from Total, mN, and mS in extremely blue $\beta_{\rm UV}$ and a break between F200W and F277W flux densities. We then attempted fitting combining two SED models with distinct ages as performed in \citet{Hashimoto+18a} and \citet{Tamura+19}: old stellar population with the Balmer break in F277W and young population with blue $\beta_{\rm UV}$ and strong nebular emission lines. This approach resulted in successful reproduction of the observed SED, as shown by the green solid spectrum and crosses in the left panel of Figure~\ref{fig:SEDres_SED_Subs}. The old and young component models are also shown by thin dashed and solid spectra, respectively.

The situation for sN1 is similar to that of sE, although the failure of the single-component fit is less severe. In the right panel of Figure~\ref{fig:SEDres_SED_Subs}, the black solid spectrum and crosses correspond to the best-fit model in the single-component fitting, which does not perfectly reproduce the blue UV slope. The two-component model (old $+$ young), shown by the red solid spectrum and crosses, yields a slightly better fit and is moderately preferred over the single-component model.

While the SEDs of the main clumps (mN, mS, and Total) are reasonably well explained by the single-component fitting as described above, the presence of underlying old stellar populations cannot be ruled out, as long as their contribution to the observed SEDs is negligible. We performed the two-components SED fitting to the main clumps to evaluate upper limits of their stellar masses. We found that the inclusion of old stellar populations, which could increase the total stellar mass by a factor of 3--10, cannot be ruled out (Section~\ref{sec:PhProp}). Such hidden old stellar populations in high-$z$ star-forming galaxies are also suggested in previous works to explain the dust formation \citep[][]{Tamura+19,Hashimoto+19a,Ren+25}. 

\subsection{Physical Properties} \label{sec:PhProp}

\begin{table*}[]
\begin{center}
\caption{Physical quantities from the multi-wavelength data. The $3\,\sigma$ upper limits are given for nondetections.} \label{tb:propAll}
\begin{tabular}{ccccccc}
\hline
\hline
  & mN &  mS &  sE &  sN1 &  sN2 & Total \\ 
\cmidrule(lr){1-7} 
\multicolumn{7}{c}{SED fitting results\tablenotemark{\rm a}} \\
Dust $A_{\rm V}$(SED) [mag] & $0.28^{+0.03}_{-0.04}$ & $0.20^{+0.02}_{-0.02}$  & $0.00^{+0.02}_{-0.00}$ & $0.00^{+0.05}_{-0.00}$ & --- & $0.20^{+0.03}_{-0.03}$ \\ 
Age [Myr] &  $5.2^{+1.7}_{-1.0}$ & $8.3^{+1.6}_{-1.6}$  & $1.4^{+0.4}_{-0.4}$ (young)  & $6.9^{+25.1}_{-5.9}$ (young) & --- & $8.3^{+2.0}_{-1.2}$ \\ 
 &  &  &  $160.9^{+207.7}_{-50.2}$ (old) & $101.5^{+18.1}_{-0.0}$ (old) &  & \\
Escape fraction $f_{\rm esc}$(SED) [\%] & $0.0^{+0.0}_{-0.0}$ & $0.0^{+8.7}_{-0.0}$ &  $0.0^{+9.7}_{-0.0}$ & $10.0^{+30.8}_{-10.0}$ & --- & $0.0^{+0.0}_{-0.0}$ \\ 
Metallicity $Z$(SED) [$Z_\odot$] & $0.40^{+0.00}_{-0.19}$ & $0.40^{+0.00}_{-0.00}$  & $0.40^{+0.47}_{-0.00}$ & $0.40^{+0.60}_{-0.20}$ & --- & $0.40^{+0.00}_{-0.18}$ \\ 
Stellar mass $M_*$ [$10^8$\,$M_\odot$] & $1.12^{+4.46}_{-0.12}$\tablenotemark{\rm b} & $1.72^{+3.54}_{-0.18}$\tablenotemark{\rm b}  & $3.22^{+1.95}_{-0.70}$ & $2.98^{+0.71}_{-0.70}$ & ---  & $6.12^{+54.52}_{-0.45}$\tablenotemark{\rm b} \\ 
SFR(SED) [$M_\odot$\,yr$^{-1}$] & $21.6^{+1.3}_{-2.2}$ & $21.5^{+1.2}_{-0.8}$ & $6.9^{+4.4}_{-0.1}$ & $2.4^{+14.6}_{-0.1}$  & ---  & $76.6^{+3.8}_{-5.1}$ \\ 
\cmidrule(lr){1-7} 
\multicolumn{7}{c}{Properties from line diagnostics} \\
Dust $A_{\rm V}$(BD) [mag]\tablenotemark{\rm c} & $0.28_{-0.28}^{+0.38}$ & $0.22_{-0.22}^{+0.35}$  & ---  & $0.0_{-0.0}^{+1.4}$  & ---  & $0.23_{-0.23}^{+0.32}$ \\ 
SFR(H$\alpha$) [$M_\odot$\,yr$^{-1}$] & $8.9 \pm 2.6$ & $9.5 \pm 2.5$  & $1.1 \pm 0.3$  & $0.9^{+1.1}_{-0.9}$  & ---  & $32.8 \pm 8.2$ \\ 
Metallicity $Z$(R3) [$Z_\odot$]\tablenotemark{\rm d} & $0.10_{-0.02}^{+0.03}$  & $0.08_{-0.02}^{+0.03}$  & ---  & $0.05_{-0.05}^{+0.23}$ & ---  & $0.07_{-0.02}^{+0.03}$ \\ 
Ionizing parameter $\log(U_{\rm ion})$ & $-2.71_{-0.08}^{+0.07}$ & $-2.78_{-0.08}^{+0.06}$  &  --- & $> -2.99$  & ---  & $-2.83 \pm 0.06$ \\ 
\cmidrule(lr){1-7} 
\multicolumn{7}{c}{Rest-frame EWs of emission lines} \\
EW$_0$(\OII\,3727d) [\AA] & $189.4 \pm 25.6$ & $162.6 \pm 18.5$ & $253.0 \pm 63.5$ & $< 94.4$ & --- &$164.1 \pm 16.2$ \\ 
EW$_0$(\NeIII\,3870) [\AA] & $81.7 \pm 13.3$ & $48.3 \pm 7.3$ & $< 62.4$ & $< 173.5$ & --- &$51.2 \pm 6.7$ \\  
EW$_0$(H$\gamma$) [\AA] & $104.9 \pm 47.2$ & $64.6 \pm 13.6$ & $< 365.9$ & $< 62.0$ & --- &$41.3 \pm 15.8$ \\ 
EW$_0$(\OIII\,4364) [\AA] & $99.2 \pm 32.5$ & $< 30.9$ & $< 76.7$ & $< 44.1$ & --- &$30.2 \pm 8.3$ \\  
EW$_0$(H$\beta$) [\AA] & $251.1 \pm 25.1$ & $165.6 \pm 16.6$ & $< 124.4$ & $66.7 \pm 76.7$ & --- &$160.5 \pm 15.9$ \\ 
EW$_0$(\OIII\,4960) [\AA] & $668.8 \pm 66.9$ & $391.9 \pm 39.2$ & $189.6 \pm 27.1$ & $141.5 \pm 23.4$ & --- &$357.5 \pm 36.9$ \\ 
EW$_0$(\OIII\,5008) [\AA] & $2016.9 \pm 201.7$ & $1229.8 \pm 123.0$ & $590.1 \pm 59.0$ & $343.8 \pm 34.4$ & --- &$1098.8 \pm 114.0$ \\
EW$_0$(H$\alpha$) [\AA] & $1580.6 \pm 158.1$ & $974.1 \pm 97.4$ & $446.5 \pm 109.8$ & $294.8 \pm 138.1$ & --- &$938.3 \pm 92.8$ \\ 
\cmidrule(lr){1-7} 
\multicolumn{7}{c}{Others} \\
$M_{\rm UV}$ [mag]\tablenotemark{\rm e} & $-19.9 \pm 0.1$ & $-20.5 \pm 0.1$  & $-18.8 \pm 0.3$  & $-19.3 \pm 0.1$  & $> -18.3$  & $-21.9_{-0.1}^{+0.2}$ \\ 
SFR(UV) [$M_\odot$\,yr$^{-1}$] & $8.6 \pm 1.5$ & $11.0 \pm 1.3$  & $0.9 \pm 0.3$  & $1.4 \pm 0.3$  & ---  & $37.4 \pm 7.0$ \\ 
$\log(\xi_{\rm ion}[\mathrm{Hz\,erg^{-1}}])$ & $25.38_{-0.18}^{+0.13}$ & $25.30_{-0.15}^{+0.11}$ & $25.45_{-0.20}^{+0.14}$ & $25.17_{-\infty}^{+0.32}$ & --- & $25.31_{-0.16}^{+0.12}$ \\ 
Escape fraction $f_{\rm esc}$(C24) [\%]\tablenotemark{\rm f} & $0.8^{+1.4}_{-0.8}$ & $2.6^{+3.4}_{-2.6}$ & --- & --- & --- & $5.2^{+8.9}_{-5.2}$ \\ 
\hline
\end{tabular}
\end{center}
\tablenotetext{\rm a}{The fitting results with the two-components model (old and young) are shown for sE and sN1, while the one-component  fits are applied for the other clumps.}
\tablenotetext{\rm b}{The $1\,\sigma$ upper errors for the stellar mass of mN, mS, and Total are given from the two-components fits to avoid underestimating the stellar mass (see text). }
\tablenotetext{\rm c}{Dust attenuation $A_{\rm V}$ from the Balmer decrement (H$\alpha/$H$\beta$) measurements. The physical quantities below in the table are estimated from the dust attenuation-corrected line fluxes. }
\tablenotetext{\rm d}{Metallicity $12 + \log(\mathrm{O/H})$ from the R3 index is estimated following \citet{Hirschmann+23}. In this table they are expressed in $Z$\,[$Z_\odot$] assuming $12 + \log(\mathrm{O/H})_\odot = 8.69$. }
\tablenotetext{\rm e}{Absolute UV magnitudes are estimated from the NIRCam F115W fluxes where no dust attenuation correction is applied. }
\tablenotetext{\rm f}{An analytic formulae of \citet{Choustikov+24a} is used to estimate the escape fractions of ionizing photons ($f_{\rm esc}$). }
\end{table*}

\begin{figure*}[t]
    \centering
    \includegraphics[width=1.0\linewidth, angle=0]{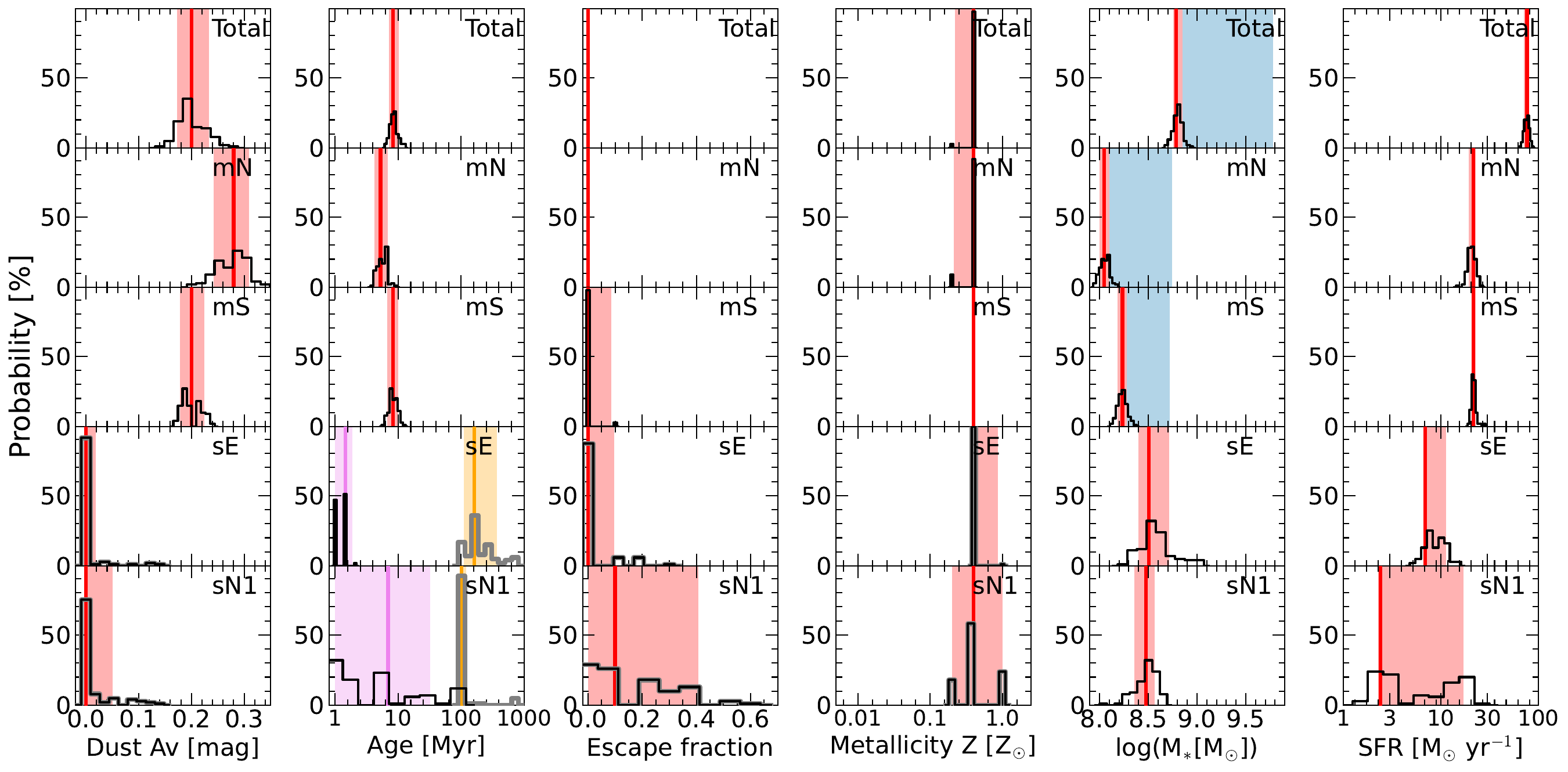}
    \caption{Probability distributions of physical quantities for the all clumps (black histogram) except for sN2. These are derrived from 100 MC iterations of the SED fitting (Section~\ref{sec:SEDfit}; \citealt{Mawatari+20,Mawatari+20b}). In each panel, vertical solid line and shade indicate the best-fit quantity and its $1\,\sigma$ uncertainty. For sE and sN1 there are two age solutions corresponding to the young and old components. In the panels of stellar mass of the main clumps (mN, mS, and Total), blue shades are also shown, which correspond to $1\,\sigma$ uncertainties taking account of hidden old stellar populations by the two-components fitting (Section~\ref{sec:SEDfit}). }
    \label{fig:SEDres_prop_all}
\end{figure*}

\begin{figure}
    \centering
    \includegraphics[width=1.0\linewidth, angle=0]{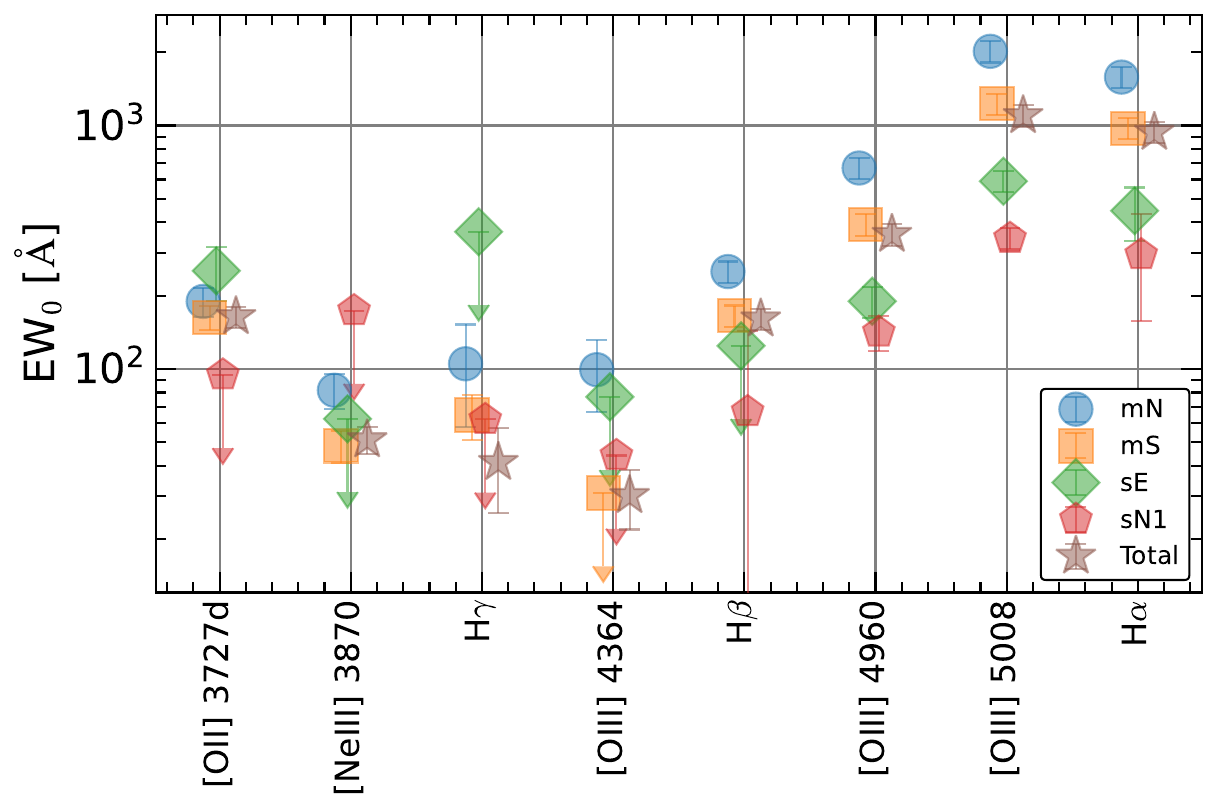}
    \caption{The rest-frame EWs of the emission lines from the multiple clumps in \tName. }
    \label{fig:LineEWs}
\end{figure}

\begin{figure*}
    \centering
    \includegraphics[width=1.0\linewidth, angle=0]{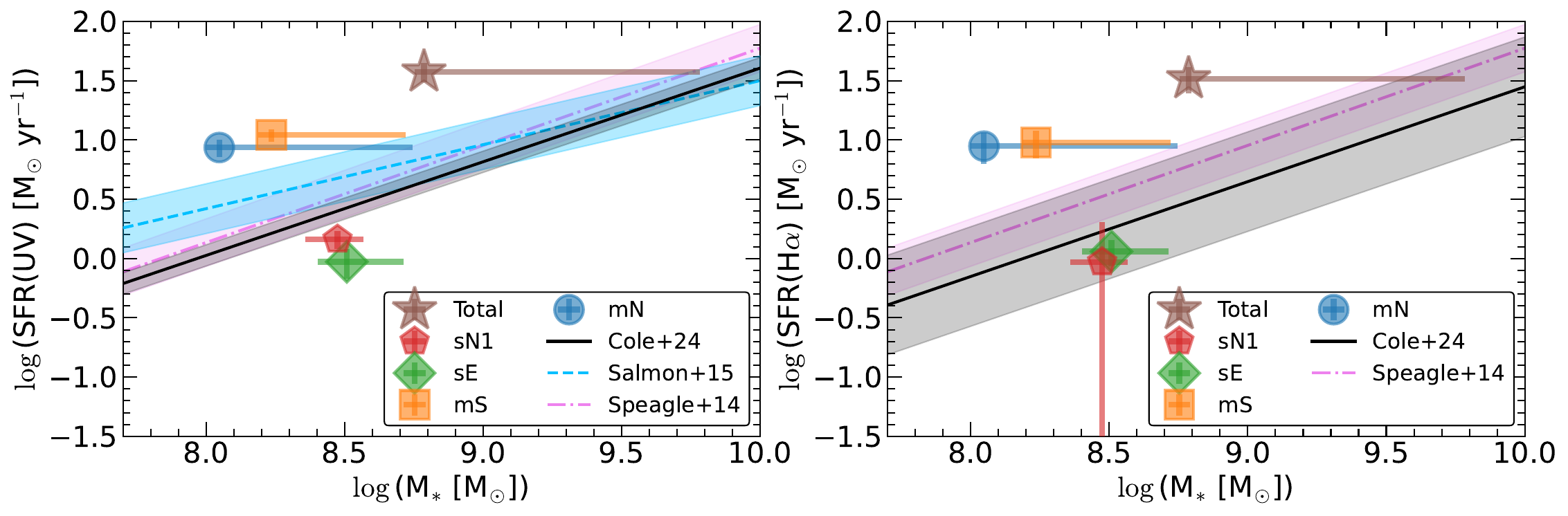}
    \caption{Stellar mass versus SFR diagram. The SFRs of \tName\ estimated from the UV and H$\alpha$ luminosities are shown in the left and right panels. The mN, mS, sN1, and Total clumps are indicated by the circle, square, pentagon and star, respectively. Star formation main-sequences from \citet{Cole+25}, \citet{Salmon+15}, and \citet{Speagle+14} are shown by solid, dashed, and dot-dashed lines, respectively. The associated shades correspond to $1\,\sigma$ uncertainty from each literature. We plot 100\,Myr- and 10\,Myr-averaged SFRs of $6 < z < 7$ galaxies from \citet{Cole+25} in comparison with our SFR(UV) and SFR(H$\alpha$). The main-sequence of \citet{Salmon+15} was measured using SFR(UV) of $z \sim 6$ galaxies. \citet{Speagle+14} compiled various previous works to establish the redshift-dependent main-sequence, from which we extract that at $z = 6.81$.  
    }
    \label{fig:SFMS}
\end{figure*}

\begin{figure}
    \centering
    \includegraphics[width=1.0\linewidth, angle=0]{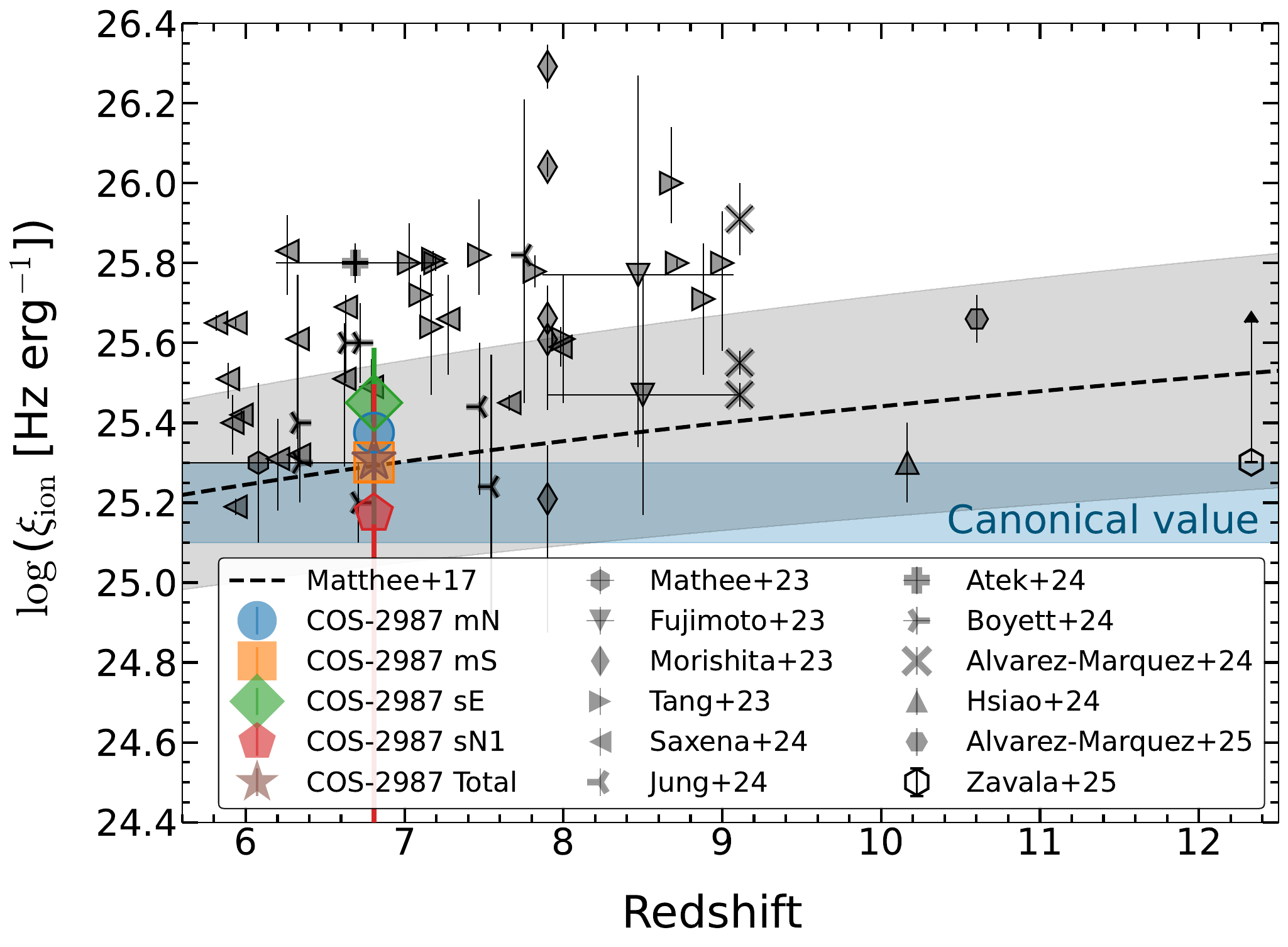}
    \caption{Ionizing photon production efficiency $\xi_{\rm ion}$ as a function of redshift. The mN, mS, sN1, and Total clumps in \tName\ are represented by the circle, square, pentagon, and star symbols, respectively. We also plot galaxies spectroscopically confirmed at $z \gtrsim 6$ with JWST from literature \citep{Fujimoto+23,Morishita+23,Matthee+23,Tang+23,Jung+24,Boyett+24a,Saxena+24,Atek+24,Alvarez-Marquez+24,Hsiao+24,Alvarez-Marquez+25,Zavala+25}. The blue shade shows the canonical value for $\xi_{\rm ion}$ \citep{Robertson+13}. The black line and shade show the redshift evolution of $\xi_{\rm ion}$ and its uncertainty \citep{Matthee+17b}. }
    \label{fig:z2xiIon}
\end{figure}

In Table~\ref{tb:propAll}, we summarize the physical quantities of individual clumps, where almost all the quantities could not be derived for sN2 due to the small numbers of detected bands and lines. The uncertainties associated with the SED fit parameters were derived by the Monte-Carlo (MC) iterations of the fitting \citep{Mawatari+20b,Mawatari+20}. Figure~\ref{fig:SEDres_prop_all} shows MC probability distribution for each clump as a function of each parameter in the SED fitting. 

The dust attenuations are well constrained as $A_{\rm V} = 0.2$--$0.3$ for mN, mS, Tot, and $A_{\rm V} \approx 0$ for sE and sN1. We also estimated $A_{\rm V}$ from the H$\alpha$ and H$\beta$ fluxes spatially integrated within each clump (Table~\ref{tb:propAll}; Section~\ref{sec:lineRatio}) to find that they are consistent with the SED fit results within the uncertainties.
The main clumps (mN, mS, and Total) have very young stellar ages of 5--10\,Myr, although the presence of older stellar populations with ages of $\gtrsim 100$\,Myr cannot be ruled out, as discussed in Section~\ref{sec:SEDfit}. In contrast, the sub-clumps (sE and sN1) require a more substantial population of $\gtrsim 100$\,Myr-old stars, such that the contributions from the old and young components become comparable in their SEDs (Figure~\ref{fig:SEDres_SED_Subs}, Table~\ref{tb:propAll}).

The balance between the young stellar ($+$nebular) component producing strong emission lines and the older stellar populations is reflected in the equivalent widths (EWs) of rest-optical emission lines \citep{LevesqueLeitherer13,TellesMelnick18}. We measured the rest-frame EW (EW$_0$) of each line of each clump using the line flux observed by NIRSpec IFS (Table~\ref{tb:fluxAll}) and the continuum flux density estimated from the best-fit model spectrum (Section~\ref{sec:SEDfit}). The EW$_0$ measurements are summarized in Table~\ref{tb:propAll} and shown in Figure~\ref{fig:LineEWs}. In the Balmer and \OIII\,5008\AA,4960\AA\ lines that are sensitive to SFR, the sub-clumps have lower EWs than the main clumps. This trend likely reflects a larger contribution from older stellar populations in the sub-clumps, diluting the emission-line EWs.

The stellar masses (M$_*$) of individual clumps range $1$--$5 \times 10^8$\,M$_{\odot}$, with large uncertainty caused by possible hidden old stellar populations for the main clumps. The SFR output by \texttt{PANHIT} corresponds to the SFR at the observed epoch, assuming a constant star formation history (SFH), which we refer to as SFR(SED). \tName, in total, has the SFR(SED) as high as 80\,M$_{\odot}$\,yr$^{-1}$ that are mainly owed by mN and mS. 
For comparison with previous works, we also estimated two types of SFRs: SFR estimated from the observed UV and H$\alpha$ luminosities ($L_{\rm UV}$ and $L_{{\rm H}\alpha}$), which we refer to SFR(UV) and SFR(H$\alpha$), respectively. The $L_{\rm UV}$ was estimated from the NIRCam F115W flux density in the same manner as in Section~\ref{sec:IRXbeta}. The $L_{\rm UV}$ and $L_{{\rm H}\alpha}$ were both corrected for the dust attenuation with the best-fit $A_{\rm V}$ from the SED fitting and from the Balmer decrement, respectively. We adopted the conversion factors from UV and H$\alpha$ luminosities to SFR(UV) and SFR(H$\alpha$) for low-metallicity galaxies following the prescriptions of \citet{Theios+19} and \citet{Shapley+23}.
It is known that the SFR(H$\alpha$) and SFR(UV) roughly trace 10\,Myr- and 100\,Myr-scale star-formation, respectively (e.g., \citealt{KennicuttEvans12,Calzetti13}). The SFR(SED) measurements are comparable to the SFR(H$\alpha$).

Figure~\ref{fig:SFMS} shows \tName\ in M$_*$ versus SFR space, where the SFR(UV) and SFR(H$\alpha$) are used in the left and right panels, respectively. We also superpose the tight relation, ``main-sequence'' of star-forming galaxies, at $6 \lesssim z \lesssim 7$ from literature \citep{Cole+25,Salmon+15,Speagle+14}. \citet{Cole+25} estimate 10\,Myr- and 100\,Myr-averaged SFRs from the Cosmic Evolution Early Release Science survey (CEERS; \citealt{Finkelstein+22,Bagley+23,Finkelstein+25}), which roughly correspond to the SFR(H$\alpha$) and SFR(UV), respectively. \citet{Salmon+15} investigate SFR(UV) using the data from the Cosmic Assembly Near-infrared Deep Extragalactic Legacy Survey (CANDELS; \citealt{Grogin+11,Koekemoer+11}). \citet{Speagle+14} compile various existing data to obtain the main-sequence as a function of the cosmic time. Ignoring the uncertainties in the M$_*$ and SFR measurements, the main clumps (mN, mS, and Total) lie about $2\,\sigma$ above the previously established main sequence in both SFR(UV) and SFR(H$\alpha$), where $\sigma$ denotes the 1$\sigma$ uncertainty of the main sequence relation. However, since the stellar masses of the main clumps could be significantly larger than their face values, we cannot definitively conclude that they are in a starburst phase relative to main-sequence galaxies. 
The sub-clumps sE and sN1 are on or below the main-sequence.

Ratio of SFR(H$\alpha$) and SFR(UV) is considered as the star-burstiness because of the difference in their timescales (e.g., \citealt{Weisz+12,Sparre+17,Atek+22,Asada+24,Endsley+25a}). 
In \tName, all the clumps have the ratios of $\sim 1$, suggesting that these clumps are not necessarily undergoing a starburst phase.

The ionizing photon production efficiency, $\xi_{\rm ion}$, is the number of ionizing photons produced per unit UV luminosity density \citep{Robertson+13}. This quantity is important to assess the ionizing budget from star-forming galaxies, particularly in the reionization era (e.g., \citealt{Robertson+13,Bouwens+16b,Matthee+17b}). 
We followed \citet{Alvarez-Marquez+24} to estimate $\xi_{\rm ion}$ of each clump from the observed H$\alpha$ luminosity ($L_{{\rm H}\alpha}$ [erg\,s$^{-1}$]) and UV luminosity density ($L_{\nu,{\rm UV}}$ [erg\,s$^{-1}$\,Hz$^{-1}$]), assuming ionizing photon escape fraction $f_{\rm esc} = 0$. The $L_{\nu,{\rm UV}}$ was estimated from the NIRCam F115W flux density. The $L_{{\rm H}\alpha}$ and $L_{\nu,{\rm UV}}$ were corrected for dust attenuation $A_{\rm V}$. We assumed the same $A_{\rm V}$ for the H$\alpha$ nebular emission and UV stellar continuum (c.f., \citealt{Erb+06c,Kashino+13}).
The estimated $\xi_{\rm ion}$ values are shown in Figure~\ref{fig:z2xiIon} along with other galaxies at $6 \lesssim z \lesssim 12$ (\citealt{Alvarez-Marquez+24} and references therein; \citealt{Hsiao+24,Alvarez-Marquez+25,Zavala+25}). The $\xi_{\rm ion}$ of mN, mS, sE, sN1, and Total distribute around the canonical value of $\log(\xi_{\rm ion} {\rm [Hz\,erg^{-1}]}) = 25.2$ \citep{Robertson+13} and comparable to the other galaxies. Our measurements are roughly consistent with the redshift evolution reported by \citet{Matthee+17b}, and also with the prediction from the FirstLight simulation ($\log(\xi_{\rm ion} {\rm [Hz,erg^{-1}]}) \sim 25.3$ at $z = 7$; \citealt{Ceverino+19}).

From the SED fitting, \tName\ prefers zero escape fraction of the ionizing photons ($f_{\rm esc} \sim 0$; Figure~\ref{fig:SEDres_prop_all}). 
Because the escape fraction is intricately related to various physical quantities, more careful estimation is needed. We used the analytic formulae of \citet{Choustikov+24a} who comprehensively investigate relations between $f_{\rm esc}$ and various observable quantities. Their formulae requires six measurements of $\beta_{\rm UV}$, dust extinction $E(B-V)$, $\log(L_{{\rm H}\beta})$, UV absolute magnitude $M_{\rm UV}$, O32 index, and R23 index $=$ (\OIII\,4960,5008\AA$+$\OII\,3727,3730\AA)$/$H$\beta$, which allows us to evaluate $f_{\rm esc}$ of mN, mS, and Total. We obtain escape fractions of $f_{\rm esc} \approx 1$--$5$\,\%, but they are not significantly different from zero given the uncertainties. (Table~\ref{tb:propAll}).

\section{Conclusions: A Comprehensive View of \tName} \label{sec:disc_view}

\begin{figure}
    \centering
    \includegraphics[width=1.0\linewidth, angle=0]{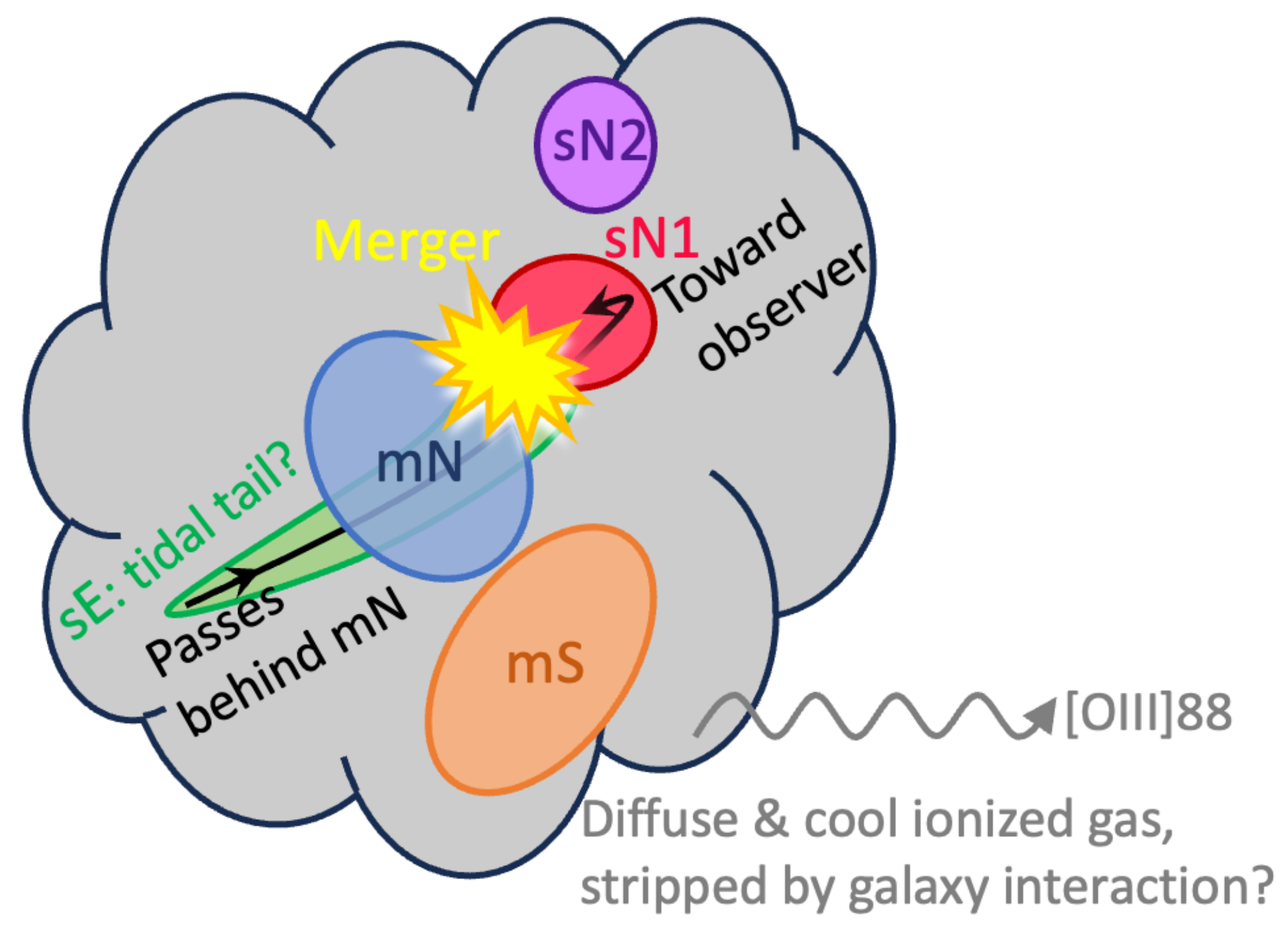}
    \caption{Schematic illustration of the physical picture of \tName. The clump sN1 is merging into mN from behind, and sE may be a tidal tail caused by the gravitational interaction. The gray shaded region corresponds to relatively low-density and low-temperature ionized gas emitting the \OIII\,88$\mu$m line, which may originate from tidal gas stripping by the merger. }
    \label{fig:SchematicPic}
\end{figure}

We here discuss the nature of \tName\ taking into account all the results in the above sections. The NIRSpec IFS data revealed the morphology and kinematics: the five clumps in the complex velocity field support the interpretation that this system is not a rotation-dominated disk (Section~\ref{sec:kinematics}). This seems inconsistent with the previous ALMA studies reporting a rotating disk \citep{Smit+18,Posses+23}. As mentioned by \citet{Posses+23}, their ALMA \CII\,158$\mu$m observations may lack the spatial resolution (the beam size $= 0\farcs44 \times 0\farcs35$) and SNR ($\sim 7$) to distinguish a rotating disk and merger, compared to the requirements proposed by \citet{Rizzo+22} (i.e., SNR $\gtrsim 10$ and more than three independent resolution elements). 

On the other hand, the difference in kinematics between our JWST and previous ALMA observations is not necessarily contradictory, because the \OIII\,5008\AA\ and \CII\,158$\mu$m line emissions trace different gas phases, warm ionized ($T \sim 10^4$\,K) and cold molecular/neutral ($T \sim 100$\,K) gas, respectively \citep{Parlanti+23,Kohandel+24,Rizzo+24,Arribas+24}.
It is plausible that, in the case of \tName, ALMA traces the underlying cold disk, whereas JWST probes the turbulent ionized gas, potentially driven by a merger or supernova feedback. 
Differences in the apparent gas kinematics depending on the emission-line tracers have also been reported in other high-$z$ galaxies. For instance, in the case of MACS1149-JD1 at $z = 9.1$, the ALMA observations of \OIII\,88$\mu$m suggested a rotating disk \citep{Tokuoka+22}, while the NIRSpec IFS of \OIII\,5008\AA\ indicated a merger \citep{Marconcini+24}. MACS0416-Y1 at $z = 8.3$ showed rotational signatures in \CII\,158$\mu$m \citep{Bakx+20}, whereas \OIII\,88$\mu$m indicated more turbulent, disordered motion \citep{Tamura+23}. Another case is COS-3018 at $z = 6.85$: \citet{Smit+18} proposed a rotating disk based on ALMA \CII\,158$\mu$m data, but \citet{Scholtz+25} later identified it as a merger based on NIRSpec IFS observations. These examples highlight the importance of carefully considering both the sensitivity and the choice of gas tracers when assessing the morpho-kinematic nature of high-$z$ galaxies.

Comparison with simulations can help the understanding of this galaxy. \citet{Nakazato+24} analyzed galaxies at $5.5 \leq z \leq 9.5$ in the FirstLight simulation \citep{Ceverino+17}. They find that about 10\,\% of the simulated galaxies show clumpy morphology in the \OIII\,5008\AA\ emission (i.e., multiple knots in $\sim 10$\,kpc extent). Interestingly, most of the \OIII\,5008\AA\ bright clumps with the line flux greater than $10^{-18}$\,erg\,s$^{-1}$\,cm$^{-2}$ are formed by major mergers enhancing the tidal compression of gas (e.g., \citealt{DiMatteo+08,Renaud+14}), not by violent disk instability accompanied with the cold gas accretion (e.g., \citealt{Dekel+09,Ceverino+10}). The similarity in the SFR surface density ($\sim 10$\,M$_\odot$\,yr$^{-1}$\,kpc$^{-2}$; Figure~\ref{fig:PhysParamMaps}) and spatial extent ($\sim 10$\,kpc) between \tName\ and the simulated clumpy galaxies suggest that the multiple clumps in \tName\ are also induced by the major merger event. The tidal tail-like structure of sE is also consistent with the merger-induced clumpy galaxies in the simulation. Thus, if \tName\ is undergoing a major merger, the five observed clumps are naturally explained as a mixture of interacting galaxies and clumps newly formed through tidal gas compression.

The physical properties of \tName\ provide further insights into its nature. The stellar masses of the clumps are comparable (mass ratio $\sim$1--3:1; Section~\ref{sec:PhProp}) and significantly higher ($\sim 10^8$--$10^9$\,M$_\odot$) than those of clumps typically formed via violent disk instability ($\sim 10^5$--$10^6$\,M$_\odot$; e.g., \citealt{Guo+15}). Such massive and comparable clumps fall within the typical definition of major mergers (e.g., \citealt{Lotz+10a,Rodriguez-Gomez+15}). Furthermore, the spatially resolved metallicity map reveals a difference in gas-phase metallicity between mN and mS, suggesting different chemical enrichment histories.

An interesting result revealed by NIRSpec high spatial resolution IFS is that the sub-clumps (sE and sN1) show very different properties compared to the main clumps: dust-free ($A_{\rm V} \approx 0$), more population of old stars with age $\gtrsim 100$\,Myr (Section~\ref{sec:PhProp}), and significantly different velocities ($\sim -40$\,km\,s$^{-1}$ for sN1; Section~\ref{sec:kinematics}).
Such diversity in clump properties has also been observed in other high-redshift merging systems \citep[e.g.,][]{Arribas+24,Sugahara+25,Jones+24a,Jones+25c,Jones+25}.
In the case of merger, sN1 may be a ``perturber'' that is originally a different galaxy and triggers the clumpy starburst in the ``host'' galaxy. The largest velocity dispersion between sN1 and mN (Section~\ref{sec:kinematics} and Appendix~\ref{ap:kinematics}) can be explained by their merger. The tail-like morphology of sE naturally evokes a tidal tail stripped from sN1 (Figure~\ref{fig:mommaps}). Based on the significant blueshift of sN1 with respect to mN, we propose a scenario in which sN1 is colliding into mN from behind, while sE represents a tidal tail that has been stripped during the interaction. This configuration is illustrated schematically in Figure~\ref{fig:SchematicPic}. In this scenario, mS and sN2 could be either originally distinct galaxies or merger-induced clumps formed within the main galaxy (mN).

Previous ground-based rest-frame UV spectroscopy detects the Ly$\alpha$ emission line from \tName, which indicates existence of an ionized bubble around this galaxy \citep{Laporte+17}. Our rest-frame optical IFS shows the high ionizing photon production efficiency (Section~\ref{sec:PhProp}). On the other hand, the ionizing photon escape fraction is estimated to be almost zero in mN, mS, and Total (Section~\ref{sec:PhProp}). A possible scenario is that the ionized bubble had already formed before the merger and the dust formed during the merger made it difficult for the ionizing photons to escape.
This idea is supported by potentially complex SFH expected from the co-existence of old and young stellar populations (Section~\ref{sec:SEDfit}).
Another scenario is that the sub-clumps with blue UV slope $\beta$ (Section~\ref{sec:IRXbeta}) actually have high escape fractions, which cannot be verified with the current data due to the lack of the required measurements (Section~\ref{sec:PhProp}; \citealt{Choustikov+24a}). We need deeper IFS observations to distinguish the scenarios.

In the companion paper \citep{Usui+25}, we investigate \tName\ focusing on the doubly-ionized oxygen lines, \OIII\,4364\AA, \OIII\,5008\AA, and \OIII\,88$\mu$m. That study suggests the presence of a multi-phase ionized gas structures, characterized by distinct combinations of electron temperatures and densities. The \OIII\,88$\mu$m line emission originates predominantly from low-temperature and low-density ionized gas. One possible origin of such diffuse and cool ionized gas is tidal stripping of gas from sN1 and mN, induced by the merger supported in this study (Figure~\ref{fig:SchematicPic}). We speculate that the dispersal of gas into extended, diffuse regions may suppress the star-burstiness of the clumps estimated from the SFR(H$\alpha$)$/$SFR(UV) ratio (Section~\ref{sec:PhProp}). Future high spatial resolution ALMA observations resolving the \OIII\,88$\mu$m morphology at a comparable scale to JWST would provide crucial insight into the physical origin and distribution of the diffuse and cooler ionized gas.


\begin{acknowledgments}

We are grateful to Nicholas Choustikov, Yuki Isobe, Nario Kuno, and Hide Yajima for useful discussion.
%
This paper makes use of the following ALMA data: ADS/JAO.ALMA \#2018.1.00429.S, \#2018.1.01551.S, \#2015.1.01111.S, and \#2018.1.01359.S. ALMA is a partnership of ESO (representing its member states), NSF (USA) and NINS (Japan), together with NRC (Canada), NSC and ASIAA (Taiwan), and KASI (Republic of Korea), in cooperation with the Republic of Chile. The Joint ALMA Observatory is operated by ESO, AUI/NRAO and NAOJ.
This work has made use of data from the European Space Agency (ESA) mission {\it Gaia} (\url{https://www.cosmos.esa.int/gaia}), processed by the {\it Gaia} Data Processing and Analysis Consortium (DPAC, \url{https://www.cosmos.esa.int/web/gaia/dpac/consortium}). Funding for the DPAC has been provided by national institutions, in particular the institutions participating in the {\it Gaia} Multilateral Agreement. The {\it Gaia} data are retrieved from the JVO portal (\url{http://jvo.nao.ac.jp/portal}) operated by the NAOJ. 
This research has made use of NASA’s Astrophysics Data System.
Data analysis was in part carried out on the Multi-wavelength Data Analysis System operated by the Astronomy Data Center (ADC), National Astronomical Observatory of Japan.
%
K.M. acknowledges the Waseda University Grant for Special Research Projects (Project number: 2025C-484) and the Japan Society for the Promotion of Science (JSPS) KAKENHI grant No. 20K14516. K.M. and A.K.I are supported by JSPS KAKENHI grant No. 23H00131. K.M. and T.H. are supported by JSPS KAKENHI Grant Number 22H01258. 
T.H. was supported by Leading Initiative for Excellent Young Researchers, MEXT, Japan (HJH02007) and by JSPS KAKENHI grant Nos. 23K22529, and 25K00020.
AKI and YS are supported by NAOJ ALMA Scientific Research Grant Numbers 2020-16B.
J.A.-M., L.C. and S.A. acknowledge support by grants PIB2021-127718NB-100 from the Spanish Ministry of Science and Innovation/State Agency of Research MCIN/AEI/10.13039/501100011033 and by “ERDF A way of making Europe”.
LC acknowledges financial support from Comunidad de Madrid under Atracci\'on de Talento grant 2018-T2/TIC-11612.
Y.N. is supported by JSPS KAKENHI Grant Number 23KJ0728 and a JSR fellowship. 
Y.W.R. was supported by JSPS KAKENHI Grant Number 23KJ2052. Y.F. acknowledges supports from JSPS KAKENHI Grant Numbers JP22K21349 and JP23K13149.
M.H. is supported by Japan Society for the Promotion of Science (JSPS) KAKENHI Grant No. 22KJ1598.
ACG acknowledges support by JWST contract B0215/JWST-GO-02926.
C.B-P acknowledge support by grant CM21\_CAB\_M2\_01 from the Program "Garant\'ia Juven\'il" from the "Comunidad de Madrid" 2021.
YT is supported by JSPS KAKENHI Grant Number 22H04939. 
MPS acknowledges funding support from the Ram\'on y Cajal program of the Spanish Ministerio de Ciencia e Innovaci\'on (RYC2021-033094-I).

\end{acknowledgments}

%

\vspace{5mm}
\facilities{JWST(STIS), ALMA}


\software{jdaviz \citep{jdaviz23}, astropy \citep{Astropy+22}, photoutils \citep{Bradley+23}, scipy \citep{Virtanen+20}, IRAF \citep{Tody86,Tody93}, SExtractor \citep{BertinArnouts96}, PANHIT \citep{Mawatari+20b}, APLPy \citep{APLpy12} }



\appendix

\section{Observed Velocity Difference and Intrinsic Velocity Dispersion} \label{ap:kinematics}

\begin{figure*}
    \centering
    \includegraphics[width=0.9\linewidth, angle=0]{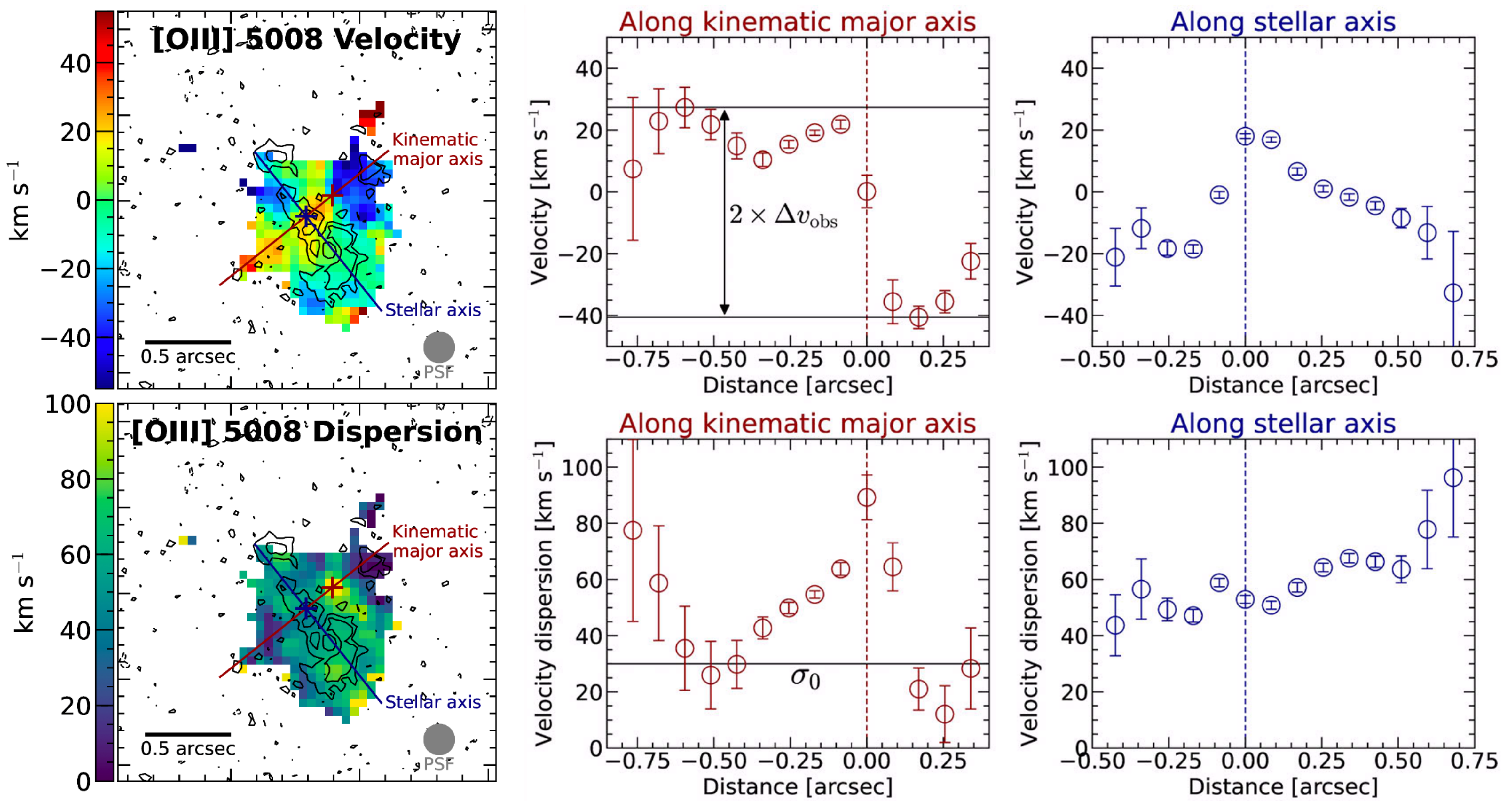}
    \caption{Top left and bottom left panels are the \OIII\,5008\AA\ velocity and velocity dispersion maps, respectively. Each panel has $2\farcs2 \times 2\farcs2$ in size. The superposed brown (navy) line and plus sign correspond to the kinematic major (stellar) axis and its center. The PSF (FWHM) at the observed \OIII\,5008\AA\ wavelength is shown by the grey filled circle at the bottom right corner of each map. Contours show the F115W flux density at $-2$ (dashed), $2$, and $4 \times \sigma$ (solid). Top middle (right) and bottom middle (right) panels show the velocity and velocity dispersion profiles along the kinematic major (stellar) axis. The vertical dashed lines correspond to the center of the axes. The horizontal solid lines in the top middle panel show the maximum and minimum velocities along the kinematic major axis, from which we calculated the observed velocity difference. The horizontal line in the bottom middle panel shows the intrinsic velocity dispersion we measured. }
    \label{fig:VelocityProfile}
\end{figure*}

To quantitatively examine the kinematic properties of \tName\ we extracted one-dimensional velocity and velocity dispersion profiles along a characteristic axis. We followed an approach applied to the KMOS$^{\rm 3D}$ IFS data \citep{Wisnioski+15}. We first determined the kinematic major axis on the two dimensional velocity map as the direction of the largest observed velocity difference. The kinematic center is defined as the spatial location of the velocity midpoint between the velocity extrema along the kinematic major axis. The kinematic major axis and center are shown in the velocity map of Figure~\ref{fig:VelocityProfile} by the brown line and cross, respectively. The kinematic major axis passes through the sE, mN, and sN1 clumps. Along the kinematic major axis we extracted one-dimensional spectra from the NIRSpec cube data with apertures equivalent to the \OIII\,5008\AA\ PSF, and performed the profile fit for the \OIII\,5008\AA\ emission line in the same manner as in Section~\ref{sec:LineSearch1}. From the line profile fit the velocity and velocity dispersion were estimated, latter of which was corrected for the instrumental spectral resolution ($\sigma_{\rm inst} \approx 48$\,km\,s$^{-1}$). The obtained velocity and velocity dispersion profiles are shown in the top middle and bottom middle panels of Figure~\ref{fig:VelocityProfile}. 

The velocity profile shows a dramatic change at the kinematic center from $v \sim +25$\,km\,s$^{-1}$ (sE and mN) to $v \sim -40$\,km\,s$^{-1}$ (sN1). The highest velocity dispersion is observed at the kinematic center. We measured an observed velocity difference and intrinsic velocity dispersion following \citet{Wisnioski+15}. The observed velocity difference is defined by $\Delta v_{\rm obs} = (v_{\rm max} - v_{\rm min}) / 2$, where $v_{\rm max}$ and $v_{\rm min}$ are the maximum and minimal velocities in the velocity profile. This quantity is equivalent to the rotation velocity in the case of a rotating edge-on disk. For \tName, we measured $\Delta v_{\rm obs} = 34$\,km\,s$^{-1}$. The intrinsic velocity dispersion, $\sigma_0$, is that measured at outer regions typically along the kinematic major axis. This quantity is ideally free of resolved motions within the galaxy and the beam smearing effect. For \tName, it is difficult to determine $\sigma_0$ from the winding velocity dispersion profile. We roughly consider $\sigma_0 \sim 30$\,km\,s$^{-1}$ (the bottom middle panel of Figure~\ref{fig:VelocityProfile}). 

We also investigated velocity and velocity dispersion profiles along the stellar mass distribution. We took an axis tracing the NIRCam F115W flux density distribution (i.e., distribution of young stars) as which we refer to the stellar axis (navy line on the velocity and velocity dispersion maps of Figure~\ref{fig:VelocityProfile}). We consider the intersection of the stellar and kinematic major axes as the stellar center for convenience. The velocity and velocity profiles measured along the stellar axis are shown in the top right and bottom right panels of Figure~\ref{fig:VelocityProfile}. We can see a velocity gradient within the northern part (negative distances; mN) and gradual decrease from northern (mN) to southern (mS) regions. The velocity dispersion gradually increases from mN to mS, which reflects the larger stellar mass in mS (Section~\ref{sec:PhProp}).

\section{PSF Matching} \label{ap:PSFmatch}

For the multiband SED analyses, we corrected the PSF difference among NIRSpec spectral data and NIRCam images. The NIRCam PSFs were generated by stacking point-sources in each band image as described in Section~\ref{sec:NIRCAMobs}. We generated NIRSpec PSF images at the wavelengths of the detected lines by averaging the cube data of the standard star 1808347 over a wavelength range within $\pm 50$\,spectral elements ($\pm 0.033\,\mu$m) centered on each line. We further rotated the NIRSpec PSF images so as to correct for the difference in instrumental positional angle between the standard star and \tName. FWHMs of the NIRCam and NIRSpec PSFs range 0\farcs06 -- 0\farcs16 and 0\farcs16 -- 0\farcs21, respectively. 

Convolution kernels were created using \texttt{PyPHER} \citep{Boucaud+16} so that all the PSFs are homogenized to the NIRSpec H$\alpha$ PSF with the largest FWHM $=$ 0\farcs21. To validate the accuracy of the kernels we first convolved the PSF images. The convolved PSFs show good agreement within 10\,\% precision for two dimensional images and within 0\farcs005 for FWHMs. 

We simply applied the kernel convolution to the eight NIRCam images. For NIRSpec we constructed PSF-matched data cube by convolving each spectral element slice image at wavelengths within $\pm 0.1\,\mu$m around the detected lines. We selected the convolution kernels whose wavelength is nearest from the given spectral element. We masked pixels whose wavelengths are more than $0.1\,\mu$m away from any lines. We again produced the line intensity maps by integrating the PSF-matched data cube over a wavelength range corresponding to $3 \times$ FWHM of each line profile measured in Section~\ref{sec:LineSearch1}. All the PSF-matched NIRCam images and NIRSpec line integrated intensity maps are shown in Figure~\ref{fig:stamps}.


\bibliographystyle{aasjournal}




\end{document}